\documentclass[12pt]{article}
\usepackage{amsmath,amssymb}
\usepackage{graphicx}
\newcommand{\eqdef}{\stackrel{\text{def}}{=}}
\newcommand{\n}{\nonumber\\}
\newcommand{\bm}{\boldsymbol}

\newcommand{\ignore}[1]{}
\numberwithin{equation}{section}
\newcommand{\Romannumeral}[1]{\uppercase\expandafter{\romannumeral#1}}
\newcommand{\I}{\text{\Romannumeral{1}}}
\newcommand{\II}{\text{\Romannumeral{2}}}
\newcommand{\III}{\text{\Romannumeral{3}}}
\newcommand{\IV}{\text{\Romannumeral{4}}}
\newtheorem{conj}{\bf Conjecture}

\allowdisplaybreaks[4]

\begin{document}

\baselineskip=20pt

\newfont{\elevenmib}{cmmib10 scaled\magstep1}
\newcommand{\preprint}{
    \begin{flushright}\normalsize \sf
     DPSU-16-2\\
   \end{flushright}}
\newcommand{\Title}[1]{{\baselineskip=26pt
   \begin{center} \Large \bf #1 \\ \ \\ \end{center}}}
\newcommand{\Author}{\begin{center}
   \large \bf Satoru Odake \end{center}}
\newcommand{\Address}{\begin{center}
     Faculty of Science, Shinshu University,\\
     Matsumoto 390-8621, Japan
   \end{center}}
\newcommand{\Accepted}[1]{\begin{center}
   {\large \sf #1}\\ \vspace{1mm}{\small \sf Accepted for Publication}
   \end{center}}

\preprint
\thispagestyle{empty}

\Title{Recurrence Relations of\\ the Multi-Indexed Orthogonal Polynomials
$\IV$ : closure relations and creation/annihilation operators}

\Author

\Address
\vspace{1cm}

\begin{abstract}
We consider the exactly solvable quantum mechanical systems whose eigenfunctions
are described by the multi-indexed orthogonal polynomials of Laguerre,
Jacobi, Wilson and Askey-Wilson types.
Corresponding to the recurrence relations with constant coefficients for the
$M$-indexed orthogonal polynomials, it is expected that the systems satisfy
the generalized closure relations.
In fact we can verify this statement for small $M$ examples.
The generalized closure relation gives the exact Heisenberg operator solution
of a certain operator, from which the creation and annihilation operators of
the system are obtained.
\end{abstract}

\section{Introduction}
\label{intro}

The exactly solvable quantum mechanical systems described by the classical
orthogonal
polynomials in the Askey scheme have two properties, shape invariance and
closure relation \cite{susyqm,os7}.
These two properties are sufficient conditions for the exact solvability.
The former leads to exact solvability in the Schr\"odinger picture and
the latter to that in the Heisenberg picture.
The closure relation gives the exact Heisenberg operator solution of the
sinusoidal coordinate and its negative/positive frequency parts provide
the creation/annihilation operators \cite{os7}.

After the pioneer works \cite{gkm08,q08}, new type of orthogonal
polynomials, exceptional/multi-indexed orthogonal polynomials, have been
studied intensively \cite{os16}--\cite{rrmiop3} (and references therein).
The exceptional orthogonal polynomials (in the wide sense)
$\{\mathcal{P}_n(\eta)|n\in\mathbb{Z}_{\geq 0}\}$ satisfy second order
differential or difference equations and form a complete set, but
there are missing degrees, by which the constraints of Bochner's theorem
and its generalizations \cite{bochner,szego} are avoided.
We distinguish the following two cases;
the set of missing degrees $\mathcal{I}=\mathbb{Z}_{\geq 0}\backslash
\{\text{deg}\,\mathcal{P}_n|n\in\mathbb{Z}_{\geq 0}\}$ is
case (1): $\mathcal{I}=\{0,1,\ldots,\ell-1\}$, or
case (2) $\mathcal{I}\neq\{0,1,\ldots,\ell-1\}$, where $\ell$ is a positive
integer. The situation of case (1) is called stable in \cite{gkm11}.
By using the multi-step Darboux transformations with appropriate seed
solutions \cite{darb,os15}, many exactly solvable deformed quantum mechanical
systems and various exceptional orthogonal polynomials with multi-indices
can be obtained.
When the virtual state wavefunctions are used as seed solutions, we obtain
case (1) and call them multi-indexed orthogonal polynomials
\cite{os25,os27,os26}.
When the eigenstate or pseudo virtual state wavefunctions are used as seed
solutions, we obtain case (2) \cite{os29,os30}.

The shape invariance of the original system is inherited by the deformed
systems for case (1), but lost for case (2) (some remnant remains
\cite{q12b,os29}).
On the other hand, the closure relation does not hold in the deformed systems.
This is because the closure relation is intimately related to the three term
recurrence relations. Roughly speaking, three terms in the r.h.s of the
closure relation \eqref{cr} correspond to three term recurrence relations.
The three term recurrence relations characterize the ordinary orthogonal
polynomial (Favard's theorem \cite{szego}).
Since the exceptional orthogonal polynomials are not ordinary orthogonal
polynomials, they do not satisfy the three term recurrence relations.
They satisfy recurrence relations with more terms
\cite{stz10}\cite{rrmiop}--\cite{rrmiop3}.
For example, the simplest exceptional orthogonal polynomials in \cite{gkm08}
satisfy five term recurrence relations \cite{stz10}.
In our previous papers \cite{rrmiop2,rrmiop3} we discussed the recurrence
relations with constant coefficients for the multi-indexed orthogonal
polynomials of Laguerre, Jacobi, Wilson and Askey-Wilson types.
Corresponding to these recurrence relations with constant coefficients,
we expect that the deformed systems satisfy the generalized closure relations
\eqref{crK}, which have more than three terms in the r.h.s.
The generalized closure relation gives the exact Heisenberg operator solution
of a certain polynomial in the sinusoidal coordinate, whose negative/positive
frequency parts provide the creation/annihilation operators.
The purpose of this paper is to expound these interesting subjects.

This paper is organized as follows.
In section \ref{sec:cr} we recapitulate the closure relation.
The closure relation is a commutator relation between the Hamiltonian and
the sinusoidal coordinate $\eta$. It gives the exact Heisenberg operator
solution of $\eta$ and its negative/positive frequency parts provide the
creation/annihilation operators.
In section \ref{sec:gcr} we generalize the closure relation.
If some function $X$ satisfies the generalized closure relation, we can
calculate the exact Heisenberg operator solution of $X$ and its
negative/positive frequency parts provide the creation/annihilation operators.
The discussion in this section is valid for general quantum mechanical systems
but the existence of the operator $X$ is assumed.
In section \ref{sec:rrr} we discuss that such operators $X$ exist in the
deformed systems described by the multi-indexed orthogonal polynomials of
Laguerre, Jacobi, Wilson and Askey-Wilson types.
The operator $X$ is a polynomial in $\eta$ giving the recurrence relations
with constant coefficients for the multi-indexed orthogonal polynomials.
The exact Heisenberg operator solution of $X$ and the creation/annihilation
operators are obtained.
Explicit examples are given in section \ref{sec:ex}.
The final section is for a summary and comments.
In Appendix\,\ref{app:diag} we present some formulas for the diagonalization
of a certain matrix.
In Appendix\,\ref{app:ex} we present more examples.

\section{Closure Relation}
\label{sec:cr}

In this section we recapitulate the closure relation and creation/annihilation
operators.

In many exactly solvable quantum mechanical models such as harmonic
oscillator, radial oscillator, Darboux-P\"oschl-Teller potential and their
discrete quantum mechanical counterparts,
the eigenfunction $\phi_n(x)$ has the following factorized form
\begin{align}
  &\mathcal{H}\phi_n(x)=\mathcal{E}_n\phi_n(x),\quad
  0=\mathcal{E}_0<\mathcal{E}_1<\mathcal{E}_2<\cdots,\\
  &\phi_n(x)=\phi_0(x)\check{P}_n(x)\ \ (n=0,1,2,\ldots),\quad
  \check{P}_n(x)\eqdef P_n\bigl(\eta(x)\bigr),
  \label{phinform}
\end{align}
where the function $\eta=\eta(x)$ is called the sinusoidal coordinate and
$P_n(\eta)$'s are classical orthogonal polynomials in $\eta$ :
Hermite, Laguerre, Jacobi, Wilson, Askey-Wilson, ($q$-)Racah, etc.
For these systems, the Hamiltonian $\mathcal{H}$ and the sinusoidal coordinate
$\eta$ satisfy the relation,
\begin{equation}
  \bigl[\mathcal{H},[\mathcal{H},\eta]\bigr]
  =\eta R_0(\mathcal{H})
  +[\mathcal{H},\eta]R_1(\mathcal{H})+R_{-1}(\mathcal{H}),
  \label{cr}
\end{equation}
where $R_i(z)$'s are polynomials in $z$,
\begin{equation}
  R_0(z)=r_0^{(2)}z^2+r_0^{(1)}z+r_0^{(0)},
  \ \ R_1(z)=r_1^{(1)}z+r_1^{(0)},
  \ \ R_{-1}(z)=r_{-1}^{(2)}z^2+r_{-1}^{(1)}z+r_{-1}^{(0)},
  \label{Ricoeff}
\end{equation}
and $r^{(j)}_i$'s are real constants ($r_0^{(2)}=r_1^{(1)}=r_{-1}^{(2)}=0$
for the ordinary QM, $\mathcal{H}=p^2+U(x)$).
This relation \eqref{cr} is called the closure relation;
The double commutator
$(\text{ad}\,\mathcal{H})^2\eta=[\mathcal{H},[\mathcal{H},\eta]]$ is expressed
as a linear combination of $(\text{ad}\,\mathcal{H})^0\,\eta=\eta$,
$(\text{ad}\,\mathcal{H})\eta=[\mathcal{H},\eta]$ and 1 with
$\mathcal{H}$-dependent coefficients $R_i(\mathcal{H})$, which are multiplied
from the right.
The closure relation implies that multiple commutators
$(\text{ad}\,\mathcal{H})^n\eta
=[\mathcal{H},[\mathcal{H},\cdots,[\mathcal{H},\eta]\cdots]]$ ($n$ times)
are also expressed as a linear combination of $\eta$, $[\mathcal{H},\eta]$
and 1 with $\mathcal{H}$-dependent coefficients,
\begin{align}
  (\text{ad}\,\mathcal{H})^n\eta
  &=\eta R_0(\mathcal{H})
  \frac{\alpha_+(\mathcal{H})^{n-1}-\alpha_-(\mathcal{H})^{n-1}}
  {\alpha_+(\mathcal{H})-\alpha_-(\mathcal{H})}
  +[\mathcal{H},\eta]
  \frac{\alpha_+(\mathcal{H})^n-\alpha_-(\mathcal{H})^n}
  {\alpha_+(\mathcal{H})-\alpha_-(\mathcal{H})}\n
  &\quad+R_{-1}(\mathcal{H})
  \frac{\alpha_+(\mathcal{H})^{n-1}-\alpha_-(\mathcal{H})^{n-1}}
  {\alpha_+(\mathcal{H})-\alpha_-(\mathcal{H})}
  -R_{-1}(\mathcal{H})R_0(\mathcal{H})^{-1}\delta_{n0},
\end{align}
where $\alpha_{\pm}(z)$ are defined by
\begin{align}
  &\alpha_{\pm}(z)\eqdef\tfrac12\bigl(R_1(z)
  \pm\sqrt{R_1(z)^2+4R_0(z)}\,\bigr),
  \label{alphapmexp}\\
  &R_1(z)=\alpha_+(z)+\alpha_-(z),\quad
  R_0(z)=-\alpha_+(z)\alpha_-(z).
  \label{R1R0}
\end{align}

{}From this, the Heisenberg operator solution of the sinusoidal coordinate
$\eta=\eta(x)$ can be obtained explicitly
\begin{align}
  e^{i\mathcal{H}t}\eta e^{-i\mathcal{H}t}
  &=\sum_{n=0}^{\infty}\frac{(it)^n}{n!}(\text{ad}\,\mathcal{H})^n\eta\n
  &=a^{(+)}e^{i\alpha_+(\mathcal{H})t}+a^{(-)}e^{i\alpha_-(\mathcal{H})t}
  -R_{-1}(\mathcal{H})R_0(\mathcal{H})^{-1},
  \label{eta(t)}
\end{align}
where $a^{(\pm)}=a^{(\pm)}(\mathcal{H},\eta)$ are
\begin{equation}
  a^{(\pm)}\eqdef\pm\Bigl([\mathcal{H},\eta]
  -\bigl(\eta+R_{-1}(\mathcal{H})R_0(\mathcal{H})^{-1}\bigr)
  \alpha_{\mp}(\mathcal{H})\Bigr)
  \bigl(\alpha_+(\mathcal{H})-\alpha_-(\mathcal{H})\bigr)^{-1}.
  \label{a^{(pm)}}
\end{equation}
Although the operators $a^{(\pm)}$ contain $\mathcal{H}$ in square roots,
they do not cause any problem when acting on the eigenfunction $\phi_n(x)$
of $\mathcal{H}$, because the operator $\mathcal{H}$ is replaced by the
eigenvalue $\mathcal{E}_n$.
For each model, we can check $R_0(\mathcal{E}_n)>0$, namely
\begin{equation}
  \alpha_+(\mathcal{E}_n)>0>\alpha_-(\mathcal{E}_n).
  \label{alphap>m}
\end{equation}
So $e^{i\alpha_+(\mathcal{H})t}$ and $e^{i\alpha_-(\mathcal{H})t}$ in
\eqref{eta(t)} are interpreted as the negative and positive frequency parts
respectively.
As for the harmonic oscillator, the coefficients of the negative and
positive frequency parts, $a^{(+)}$ and $a^{(-)}$, provide the creation and
annihilation operators respectively.

Since $P_n(\eta)$ are orthogonal polynomials, they satisfy the three term
recurrence relations (we set $P_n(\eta)=0$ for $n<0$),
\begin{equation}
  \eta P_n(\eta)=A_nP_{n+1}(\eta)+B_nP_n(\eta)+C_nP_{n-1}(\eta)
  \ \ (n\geq 0).
\end{equation}
This and \eqref{phinform} imply
\begin{equation}
  \eta\phi_n(x)=A_n\phi_{n+1}(x)+B_n\phi_n(x)+C_n\phi_{n-1}(x)
  \ \ (n\geq 0).
\end{equation}
Action of \eqref{eta(t)} on $\phi_n(x)$ is
\begin{equation*}
  e^{i\mathcal{H}t}\eta e^{-i\mathcal{H}t}\phi_n(x)
  =e^{i\alpha_+(\mathcal{E}_n)t}a^{(+)}\phi_n(x)
  +e^{i\alpha_-(\mathcal{E}_n)t}a^{(-)}\phi_n(x)
  -R_{-1}(\mathcal{E}_n)R_0(\mathcal{E}_n)^{-1}\phi_n(x).
\end{equation*}
On the other hand the l.h.s. turns out to be
\begin{align*}
  e^{i\mathcal{H}t}\eta e^{-i\mathcal{H}t}\phi_n(x)
  &=e^{i\mathcal{H}t}\eta e^{-i\mathcal{E}_nt}\phi_n(x)\n
  &=e^{-i\mathcal{E}_nt}e^{i\mathcal{H}t}\bigl(
  A_n\phi_{n+1}(x)+B_n\phi_n(x)+C_n\phi_{n-1}(x)\bigr)\n
  &=e^{i(\mathcal{E}_{n+1}-\mathcal{E}_n)t}A_n\phi_{n+1}(x)
  +B_n\phi_n(x)
  +e^{i(\mathcal{E}_{n-1}-\mathcal{E}_n)t}C_n\phi_{n-1}(x).
\end{align*}
Comparing these $t$-dependence (see \eqref{alphap>m}), we obtain
\begin{align}
  &\alpha_{\pm}(\mathcal{E}_n)=\mathcal{E}_{n\pm 1}-\mathcal{E}_n,\\
  &a^{(+)}\phi_n(x)=A_n\phi_{n+1}(x),\quad
  a^{(-)}\phi_n(x)=C_n\phi_{n-1}(x),\\
  &-R_{-1}(\mathcal{E}_n)R_0(\mathcal{E}_n)^{-1}=B_n.
\end{align}
Therefore $a^{(+)}$ and $a^{(-)}$ are creation and annihilation operators,
respectively. (The normalizability of $a^{(+)}\phi_n(x)$ depends on the model
and $n$.)

The creation and annihilation operators for eigenpolynomials are obtained
by the similarity transformation in terms of the groundstate wavefunction
$\phi_0(x)$.
The similarity transformed Hamiltonian is
\begin{equation}
  \widetilde{\mathcal{H}}\eqdef
  \phi_0(x)^{-1}\circ\mathcal{H}\circ\phi_0(x),\quad  
  \widetilde{\mathcal{H}}\check{P}_n(x)=\mathcal{E}_n\check{P}_n(x).
\end{equation}
Corresponding to \eqref{cr}, this also satisfies the closure relation
\begin{equation}
  \bigl[\widetilde{\mathcal{H}},[\widetilde{\mathcal{H}},\eta]\bigr]
  =\eta R_0(\widetilde{\mathcal{H}})
  +[\widetilde{\mathcal{H}},\eta]R_1(\widetilde{\mathcal{H}})
  +R_{-1}(\widetilde{\mathcal{H}}).
\end{equation}
{}From the creation and annihilation operators
$a^{(\pm)}=a^{(\pm)}(\mathcal{H},\eta)$, we obtain the creation and
annihilation operators for eigenpolynomials,
\begin{align}
  &\tilde{a}^{(\pm)}\eqdef\phi_0(x)^{-1}\circ a^{(\pm)}(\mathcal{H},\eta)
  \circ\phi_0(x)=a^{(\pm)}(\widetilde{\mathcal{H}},\eta),\\
  &\tilde{a}^{(+)}\check{P}_n(x)=A_n\check{P}_{n+1}(x),\quad
  \tilde{a}^{(-)}\check{P}_n(x)=C_n\check{P}_{n-1}(x).
\end{align}

The normalization constants of the orthogonality relations, $h_n$,
\begin{equation}
  (\phi_n,\phi_m)=h_n\delta_{nm},\quad
  (f,g)\eqdef\int_{x_1}^{x_2}\!\!dx\,f^*(x)g(x)dx,
  \label{inner}
\end{equation}
are related to the coefficients of the three term recurrence relations.
By calculating $(\phi_n,\eta\phi_m)=(\eta\phi_n,\phi_m)$ as
\begin{equation*}
  h_n(A_m\delta_{n,m+1}+B_m\delta_{n,m}+C_m\delta_{n,m-1})
  =h_m(A_n\delta_{n+1,m}+B_n\delta_{n,m}+C_n\delta_{n-1,m}),
\end{equation*}
we obtain
\begin{equation}
  A_nh_{n+1}=C_{n+1}h_n\ \ (n\geq 0).
  \label{Anhn+1=}
\end{equation}

\section{Generalization}
\label{sec:gcr}

In this section we generalize the closure relation.
We consider general quantum mechanical systems with the Hamiltonian
$\mathcal{H}$ and the coordinate $x$.

Let us consider an operator $X$, which is a function of $\eta=\eta(x)$,
\begin{equation}
  X=X(\eta)=X\bigl(\eta(x)\bigr)\eqdef\check{X}(x).
\end{equation}
We assume that the $K$-times commutator of $\mathcal{H}$ and $X$ has the
following form:
\begin{equation}
  (\text{ad}\,\mathcal{H})^KX
  =\sum_{i=0}^{K-1}(\text{ad}\,\mathcal{H})^iX\cdot R_i(\mathcal{H})
  +R_{-1}(\mathcal{H}).
  \label{crK}
\end{equation}
Here $R_i(z)=R^X_i(z)$ is a polynomial in $z$ and their coefficients are
real numbers depending on the choice of $X$.
We call this the closure relation of order $K$.
The closure relation reviewed in \S\,\ref{sec:cr} corresponds to $K=2$.

The closure relation of order $K$ \eqref{crK} implies that the multiple
commutation relation of $\mathcal{H}$ and $X$, $(\text{ad}\,\mathcal{H})^nX$,
can be expressed as a linear combination of $(\text{ad}\,\mathcal{H})^iX$
($0\leq i\leq K-1$) and 1 with $\mathcal{H}$-dependent coefficients,
which are multiplied from the right,
\begin{equation}
  (\text{ad}\,\mathcal{H})^nX
  =\sum_{i=0}^{K-1}(\text{ad}\,\mathcal{H})^iX\cdot R_i^{[n]}(\mathcal{H})
  +R_{-1}^{[n]}(\mathcal{H})\ \ (n\geq 0).
  \label{adHnX}
\end{equation}
The initial conditions of $R_i^{[n]}(z)$ are
\begin{equation}
  R_i^{[n]}(z)=\left\{
  \begin{array}{ll}
  \delta_{ni}&(0\leq n\leq K-1\,;-1\leq i\leq K-1)\\[2pt]
  R_i(z)&(n=K\,;-1\leq i\leq K-1)
  \end{array}\right..
  \label{Rinit}
\end{equation}
By applying $\text{ad}\,\mathcal{H}$ to \eqref{adHnX},
\begin{align*}
  &\quad(\text{ad}\,\mathcal{H})^{n+1}X
  =(\text{ad}\,\mathcal{H})(\text{ad}\,\mathcal{H})^nX\n
  &=(\text{ad}\,\mathcal{H})\Bigl(
  (\text{ad}\,\mathcal{H})^{K-1}X\cdot R^{[n]}_{K-1}(\mathcal{H})
  +\sum_{i=0}^{K-2}(\text{ad}\,\mathcal{H})^iX\cdot R_i^{[n]}(\mathcal{H})
  +R_{-1}^{[n]}(\mathcal{H})\Bigr)\n
  &=\Bigl(\sum_{i=0}^{K-1}(\text{ad}\,\mathcal{H})^iX\cdot
  R_i(\mathcal{H})+R_{-1}(\mathcal{H})\Bigr)\cdot
  R^{[n]}_{K-1}(\mathcal{H})
  +\sum_{i=1}^{K-1}(\text{ad}\,\mathcal{H})^iX\cdot
  R_{i-1}^{[n]}(\mathcal{H}),
\end{align*}
the recurrence relations for $R_i^{[n]}(z)$ are obtained:
\begin{equation}
  R_i^{[n+1]}(z)=R_i(z)R_{K-1}^{[n]}(z)+\theta(1\leq i\leq K-1)
  R_{i-1}^{[n]}(z)\ \ (n\geq 0\,;-1\leq i\leq K-1),
  \label{Rrec}
\end{equation}
where $\theta(\text{Prop})$ is a step function for a proposition,
$\theta(\text{True})=1$ and $\theta(\text{False})=0$.
By introducing a matrix $A$ and a vector $\vec{R}^{[n]}$
\begin{equation}
  A\eqdef\begin{pmatrix}
  0&&&&&R_0\\
  1&0&&\text{\Large$O$}&&R_1\\
  &1&0&&&R_2\\
  &&\ddots&\ddots&&\vdots\\
  &\text{\Large$O$}&&1&0&R_{K-2}\\
  &&&&1&R_{K-1}
  \end{pmatrix},\quad
  \vec{R}^{[n]}\eqdef\begin{pmatrix}
  R^{[n]}_0\\
  R^{[n]}_1\\
  R^{[n]}_2\\
  \vdots\\
  R^{[n]}_{K-1}
  \end{pmatrix},
  \label{matA,vecR}
\end{equation}
this recurrence relations \eqref{Rrec} with the initial conditions
\eqref{Rinit} can be rewritten as
\begin{align}
  &\vec{R}^{[n+1]}=A\vec{R}^{[n]}\ \ (n\geq 0),\quad
  \vec{R}^{[0]}={}^t(1\ 0\ 0\ \cdots\ 0),\\
  &R_{-1}^{[n+1]}=R_{-1}R_{K-1}^{[n]}\ \ (n\geq 0),\quad
  R_{-1}^{[0]}=0.
\end{align}
These are easily solved as
\begin{align}
  &\vec{R}^{[n]}=A^n\vec{R}^{[0]}\ \ (n\geq 0),\\
  &R_{-1}^{[n]}=\theta(n\geq 1)R_{-1}R_{K-1}^{[n-1]}
  =\theta(n\geq 1)R_{-1}\cdot\bigl(A^{n-1}\vec{R}^{[0]}\bigr)_K
  \quad(n\geq 0).
\end{align}

By using the properties of the matrix $A$ given in Appendix \ref{app:diag},
let us calculate $A^n$.
Although the matrix elements of $A$ depend on $\mathcal{H}$, we can use the
formulas in Appendix \ref{app:diag}, because the operator appearing in this
calculation is $\mathcal{H}$ only.
We assume that the matrix $A$ has $K$ distinct real non-vanishing eigenvalues
$\alpha_i=\alpha_i(\mathcal{H})$ and they are indexed in decreasing order
\begin{equation}
  \alpha_i(z)\neq 0,\quad
  \alpha_1(z)>\alpha_2(z)>\cdots>\alpha_K(z)\ \ (z\geq 0).
  \label{alpha}
\end{equation}
We note that $R_i$ ($0\leq i\leq K-1$) is expressed by $\alpha_j$,
\begin{equation}
  R_i=(-1)^{K-i-1}\!\!\!\!\!\!\!\!\!\!\!
  \sum_{1\leq j_1<j_2<\cdots<j_{K-i}\leq K}\!\!\!\!\!\!\!\!\!\!\!\!\!\!
  \alpha_{j_1}\alpha_{j_2}\cdots\alpha_{j_{K-i}}
  \ \ (0\leq i\leq K-1).
\end{equation}
The eigenvector $\vec{p}_j=(p_{ij})_{1\leq i\leq K}$
($p_{ij}=p_{ij}(\mathcal{H})$) corresponding to the eigenvalue $\alpha_j$ has
a simple expression
\begin{equation}
  p_{ij}\eqdef\alpha_j^{K-i}-\sum_{k=1}^{K-i}R_{K-k}\,\alpha_j^{K-i-k}.
\end{equation}
Note that $p_{Kj}=1$.
Since the matrix $A$ is diagonalized by the matrix
$P=(p_{ij})_{1\leq i,j\leq K}
=\bigl(\vec{p}_1\ \vec{p}_2\ \cdots\ \vec{p}_K\bigr)$,
$P^{-1}AP=\text{diag}(\alpha_1,\alpha_2,\ldots,\alpha_K)$, we obtain
\begin{equation}
  A^n=P\,\text{diag}(\alpha_1^n,\alpha_2^n,\ldots,\alpha_K^n)P^{-1}.
\end{equation}
The matrix element $(P^{-1})_{ji}$ and the sum
$\sum_{j=1}^K\alpha_j^{-1}(P^{-1})_{j1}$ are
\begin{equation}
  (P^{-1})_{ji}=\alpha_j^{i-1}
  \prod_{\genfrac{}{}{0pt}{}{k=1}{k\neq j}}^K(\alpha_j-\alpha_k)^{-1},\quad
  \sum_{j=1}^K\alpha_j^{-1}(P^{-1})_{j1}=R_0^{-1}.
  \label{Pinvj1}
\end{equation}

Next let us calculate the Heisenberg operator
$e^{it\mathcal{H}}Xe^{-it\mathcal{H}}$.
By defining $\vec{X}$,
\begin{equation}
  \vec{X}\eqdef\begin{pmatrix}
  X\\
  (\text{ad}\,\mathcal{H})X\\
  (\text{ad}\,\mathcal{H})^2X\\
  \vdots\\
  (\text{ad}\,\mathcal{H})^{K-1}X
  \end{pmatrix},
\end{equation}
eq.\,\eqref{adHnX} gives
\begin{equation}
  (\text{ad}\,\mathcal{H})^nX
  ={}^t\!\vec{X}\vec{R}^{[n]}+R_{-1}^{[n]}
  ={}^t\!\vec{X}A^n\vec{R}^{[0]}
  +\theta(n\geq 1)R_{-1}\cdot\bigl(A^{n-1}\vec{R}^{[0]}\bigr)_K
  \quad(n\geq 0).
\end{equation}
Then we have
\begin{align}
  &\quad e^{it\mathcal{H}}Xe^{-it\mathcal{H}}
  =\sum_{n=0}^{\infty}\frac{(it)^n}{n!}(\text{ad}\,\mathcal{H})^nX\n
  &=\sum_{n=0}^{\infty}\frac{(it)^n}{n!}\Bigl(
  {}^t\!\vec{X}A^n\vec{R}^{[0]}
  +\theta(n\geq 1)R_{-1}\cdot\bigl(A^{n-1}\vec{R}^{[0]}\bigr)_K\Bigr)\n
  &={}^t\!\vec{X}e^{itA}\vec{R}^{[0]}
  +iR_{-1}\int_0^t\!ds\,\bigl(e^{isA}\vec{R}^{[0]}\bigr)_K\n
  &={}^t\!\vec{X}P
  \cdot\text{diag}(e^{i\alpha_1t}\cdots e^{i\alpha_Kt})\cdot
  P^{-1}\vec{R}^{[0]}
  +iR_{-1}\int_0^t\!ds\,\Bigl(P
  \cdot\text{diag}(e^{i\alpha_1s}\cdots e^{i\alpha_Ks})\cdot
  P^{-1}\vec{R}^{[0]}\Bigr)_K\n
  &=\sum_{j=1}^K{}^t\!\vec{X}\vec{p}_je^{i\alpha_jt}(P^{-1})_{j1}
  +iR_{-1}\int_0^t\!ds\sum_{j=1}^Kp_{Kj}e^{i\alpha_js}(P^{-1})_{j1}\n
  &=\sum_{j=1}^K{}^t\!\vec{X}\vec{p}_je^{i\alpha_jt}(P^{-1})_{j1}
  +R_{-1}\sum_{j=1}^K\Bigl[\alpha_j^{-1}e^{i\alpha_js}(P^{-1})_{j1}
  \Bigr]_0^t\n
  &=\sum_{j=1}^K\bigl({}^t\!\vec{X}\vec{p}_j
  +R_{-1}\alpha_j^{-1}\bigr)(P^{-1})_{j1}e^{i\alpha_jt}
  -R_{-1}\sum_{j=1}^K\alpha_j^{-1}(P^{-1})_{j1}\n
  &=\sum_{j=1}^Ka^{(j)}e^{i\alpha_jt}-R_{-1}R_0^{-1}.
  \label{X(t)}
\end{align}
In the last line we have defined $a^{(j)}=a^{(j)}(\mathcal{H},X)$
($1\leq j\leq K$) as
\begin{equation}
  a^{(j)}\eqdef\bigl({}^t\!\vec{X}\vec{p}_j
  +R_{-1}\alpha_j^{-1}\bigr)(P^{-1})_{j1}.
  \label{a(j)}
\end{equation}
{\bf Remark}$\,$ Since the operation $\text{ad}\,\mathcal{H}$ is a
derivation, the closure relation of order $K$ \eqref{crK} is interpreted as
a differential equation. More explicitly, it is stated as follows.
Since a Heisenberg operator $F_{\text{H}}(t)$ satisfies the Heisenberg
equation $i\frac{d}{dt}F_{\text{H}}(t)=[F_{\text{H}}(t),H]$, the operation
$\text{ad}\,\mathcal{H}$ for Heisenberg operators is a derivative
$-i\frac{d}{dt}$.
Then the closure relation of order $K$ \eqref{crK} means a differential
equation for the Heisenberg operator
$X_{\text{H}}(t)=e^{it\mathcal{H}}Xe^{-it\mathcal{H}}$,
$(-i)^K\frac{d^KX_{\text{H}}(t)}{dt^K}
-\sum\limits_{l=0}^{K-1}(-i)^l\frac{d^lX_{\text{H}}(t)}{dt^l}R_l(\mathcal{H})
=R_{-1}(\mathcal{H})$,
which is a $K$-th order linear differential equation with `constant'
coefficients $R_l(\mathcal{H})$ ($0\leq l\leq K-1$) and a `constant'
inhomogeneous term $R_{-1}(\mathcal{H})$.
Such a $K$-th order differential equation is converted into a coupled first
order linear differential equation, which is expressed neatly in vector
and matrix notation, see the matrix $A$ \eqref{matA,vecR}.
Its general solution is a linear combination of $e^{i\alpha_jt}$ (which
is a general solution of the homogeneous equation) plus a particular solution
(which corresponds to an inhomogeneous term), see the last line of \eqref{X(t)}.

Although these operators $a^{(j)}$ \eqref{a(j)} contain $\mathcal{H}$ in a
complicated way, they do not cause any problem when acting on the
eigenfunction of $\mathcal{H}$, because the operator $\mathcal{H}$ is
replaced by the eigenvalue.
By noting
\begin{align}
  [\mathcal{H},{}^t\!\vec{X}]
  &=\bigl((\text{ad}\,\mathcal{H})X,(\text{ad}\,\mathcal{H})^2X,\cdots,
  (\text{ad}\,\mathcal{H})^{K-1}X,(\text{ad}\,\mathcal{H})^KX\bigr)\n
  &=\bigl((\text{ad}\,\mathcal{H})X,(\text{ad}\,\mathcal{H})^2X,\cdots,
  (\text{ad}\,\mathcal{H})^{K-1}X,
  \sum_{i=0}^{K-1}(\text{ad}\,\mathcal{H})^{i-1}X\cdot R_i+R_{-1}\bigr)\n
  &={}^t\!\vec{X}A+(0,0,\cdots,0,R_{-1}),
\end{align}
we obtain
\begin{align}
  [\mathcal{H},a^{(j)}]
  &=[\mathcal{H},{}^t\!\vec{X}]\vec{p}_j(P^{-1})_{j1}
  =\bigl({}^t\!\vec{X}A+(0,0,\cdots,0,R_{-1})\bigr)\vec{p}_j(P^{-1})_{j1}\n
  &=\bigl({}^t\!\vec{X}\alpha_j\vec{p}_j+R_{-1}p_{Kj})\bigr)(P^{-1})_{j1}
  =a^{(j)}\alpha_j.
\end{align}
This relation implies
\begin{align}
  \mathcal{H}\psi(x)=\mathcal{E}\psi(x)
  \ \ \Rightarrow\ \ \mathcal{H}a^{(j)}\psi(x)
  &=a^{(j)}\bigl(\mathcal{H}+\alpha_j(\mathcal{H})\bigr)\psi(x)
  =a^{(j)}\bigl(\mathcal{E}+\alpha_j(\mathcal{E})\bigr)\psi(x)\n
  &=\bigr(\mathcal{E}+\alpha_j(\mathcal{E})\bigr)a^{(j)}\psi(x).
\end{align}
Therefore the operator $a^{(j)}$ maps a solution of the Schr\"odinger
equation with energy $\mathcal{E}$ to that with energy
$\mathcal{E}+\alpha_j(\mathcal{E})$.
Namely it is a creation operator ($\alpha_j(\mathcal{E})>0$) or an
annihilation operator ($\alpha_j(\mathcal{E})<0$)
(we have assumed $\psi(x)$ and $a^{(j)}\psi(x)$ are normalizable).

The discussion given in this section is very general but we have assumed
the existence of $X$ satisfying the generalized closure relation \eqref{crK}.
In the next section we discuss that such $X$'s really exist in the systems
described by the multi-indexed orthogonal polynomials.

\section{Relation to the Recurrence Relations}
\label{sec:rrr}

In this section we consider exactly solvable systems described by the
multi-indexed orthogonal polynomials of Laguerre(L), Jacobi(J), Wilson(W)
and Askey-Wilson(AW) types.
We discuss the existence of the generalized closure relations together with
their connection to the recurrence relations with constant coefficients for the
multi-indexed orthogonal polynomials.
We follow the notation in \cite{rrmiop,rrmiop2,rrmiop3}.

\subsection{Multi-indexed orthogonal polynomials}
\label{sec:miop}

Isospectral deformations of the exactly solvable systems described by
L, J, W and AW polynomials are obtained by the $M$-step Darboux
transformations with the virtual state wavefunctions as seed solutions.
The deformed systems are labeled by
$\mathcal{D}=\{d_1,\ldots,d_M\}=\{d^{\I}_1,\ldots,d^{\I}_{M_{\I}},
d^{\II}_1,\ldots,d^{\II}_{M_{\II}}\}$ ($M=M_{\I}+M_{\II}$),
which are the degrees and types of the virtual state wavefunctions, and
their eigenstates have the following form,
\begin{align}
  &\mathcal{H}_{\mathcal{D}}\phi_{\mathcal{D}\,n}(x)
  =\mathcal{E}_n\phi_{\mathcal{D}\,n}(x),\quad
  0=\mathcal{E}_0<\mathcal{E}_1<\mathcal{E}_2<\cdots,\\
  &\phi_{\mathcal{D}\,n}(x)=\Psi_{\mathcal{D}}(x)
  \check{P}_{\mathcal{D},n}(x)\ \ (n=0,1,2,\ldots),\quad
  \check{P}_{\mathcal{D},n}(x)\eqdef P_{\mathcal{D},n}\bigl(\eta(x)\bigr).
\end{align}
Here $P_{\mathcal{D},n}(\eta)$'s are multi-indexed orthogonal polynomials.
The Hamiltonian $\mathcal{H}_{\mathcal{D}}$ of the deformed system is the
second order differential or difference operator,
\begin{align}
  \text{L,\,J}:&\ \ \mathcal{H}_{\mathcal{D}}=p^2+U_{\mathcal{D}}(x),
  \label{HoQM}\\
  \text{W,\,AW}:&\ \ \mathcal{H}_{\mathcal{D}}
  =\sqrt{V_{\mathcal{D}}(x)V^*_{\mathcal{D}}(x-i\gamma)}\,e^{\gamma p}
  +\sqrt{V^*_{\mathcal{D}}(x)V_{\mathcal{D}}(x+i\gamma)}\,e^{-\gamma p}
  -V_{\mathcal{D}}(x)-V^*_{\mathcal{D}}(x),
  \label{HidQM}
\end{align}
where $p=-i\frac{d}{dx}$ and $\gamma=1$ for W, $\gamma=\log q$ for AW.
The deformed potential $U_{\mathcal{D}}(x)$ and potential function
$V_{\mathcal{D}}(x)$ are expressed in terms of the original $U(x)$, $V(x)$
and the denominator polynomial $\Xi_{\mathcal{D}}(\eta)$.
The explicit forms of $P_{\mathcal{D},n}(\eta)$, $\Xi_{\mathcal{D}}(\eta)$,
$\Psi_{\mathcal{D}}(x)$, $U_{\mathcal{D}}(x)$, $V_{\mathcal{D}}(x)$, etc.
can be found in \cite{rrmiop,rrmiop2,rrmiop3}.

Since the multi-indexed orthogonal polynomials are not the ordinary orthogonal
polynomials, they do not satisfy the three term recurrence relations.
They satisfy the recurrence relations with more terms; $3+2M$ term recurrence
relations with variable dependent coefficients \cite{rrmiop}.
It is conjectured that they also satisfy  $1+2L$ term
recurrence relations with constant coefficients \cite{rrmiop2,rrmiop3}
of the form,
\begin{equation}
  X(\eta)P_{\mathcal{D},n}(\eta)
  =\sum_{k=-L}^Lr_{n,k}^{X,\mathcal{D}}P_{\mathcal{D},n+k}(\eta)
  \ \ (\forall n\geq 0),
  \label{XP}
\end{equation}
in which $X(\eta)$ is a degree $L$ polynomial in $\eta$ (with real number
coefficients) and $r_{n,k}^{X,\mathcal{D}}$'s are constants.
We have set $P_{\mathcal{D},n}(\eta)=0$ for $n<0$.
The polynomial $X(\eta)$ \eqref{XP} depends on the denominator polynomial
$\Xi_{\mathcal{D}}(\eta)$ \cite{rrmiop2}:
\begin{equation}
  X(\eta)=\left\{
  \begin{array}{ll}
  \int_0^{\eta}\Xi_{\mathcal{D}}(y)Y(y)dy&:\text{L,\,J}\\[4pt]
  I[\Xi_{\mathcal{D}}Y](\eta)&:\text{W,\,AW}
  \end{array}\right.,\quad
  \text{deg}\,X(\eta)=L=\ell_{\mathcal{D}}+\text{deg}\,Y(\eta)+1,
  \label{X}
\end{equation}
where $Y(\eta)$ is an arbitrary polynomial in $\eta$ and the map $I[\cdot]$
is given in \cite{rrmiop2} and $\ell_{\mathcal{D}}$ is
\begin{equation}
  \ell_{\mathcal{D}}=\sum_{j=1}^Md_j-\frac12M(M-1)+2M_{\I}M_{\II}.
  \label{lD}
\end{equation}
So long as the two polynomials in $\eta$,
$\Xi_{\mathcal{D}}(\eta)=\Xi_{d_1\ldots d_M}(\eta)$ and
$\Xi_{d_1\ldots d_{M-1}}(\eta)$ (with some modification for W and AW),
do not have common roots, the above form exhausts all possible $X(\eta)$
giving rise to the recurrence relations with constant coefficients \eqref{XP}.
This conjecture is proved for L and J in \cite{rrmiop3}.

The normalization constants of the orthogonality relations,
$h_{\mathcal{D}\,n}$,
are related to those of the original system \eqref{inner},
\begin{equation}
  (\phi_{\mathcal{D}\,n},\phi_{\mathcal{D}\,m})
  =h_{\mathcal{D}\,n}\delta_{nm},\quad
  h_{\mathcal{D}\,n}=h_n\prod_{j=1}^M(\mathcal{E}_n-\tilde{\mathcal{E}}_{d_j}),
  \label{hDn}
\end{equation}
where $\tilde{\mathcal{E}}_{d_j}$'s are energies of the virtual states,
see \eqref{dataL}, \eqref{dataJ}, \eqref{dataW} and \eqref{dataAW}.
As in \eqref{Anhn+1=}, the normalization constants and the coefficients of
the recurrence relations are related.
Since the recurrence relations \eqref{XP} give
\begin{equation}
  X(\eta)\phi_{\mathcal{D}\,n}(x)
  =\sum_{k=-L}^Lr_{n,k}^{X,\mathcal{D}}\phi_{\mathcal{D}\,n+k}(x)
  \ \ (n\geq 0),
  \label{Xphi}
\end{equation}
we have
\begin{align*}
  &\quad\ (\phi_{\mathcal{D}\,n},X\phi_{\mathcal{D}\,n-l})
  =\theta(-L\leq l\leq L)r_{n-l,l}^{X,\mathcal{D}}\,h_{\mathcal{D}\,n}\\
  &=(X\phi_{\mathcal{D}\,n},\phi_{\mathcal{D}\,n-l})
  =\theta(-L\leq l\leq L)r_{n,-l}^{X,\mathcal{D}}\,h_{\mathcal{D}\,n-l},
\end{align*}
which means
\begin{equation}
  r_{n-l,l}^{X,\mathcal{D}}\,h_{\mathcal{D}\,n}
  =r_{n,-l}^{X,\mathcal{D}}\,h_{\mathcal{D}\,n-l}
  \ \ (-L\leq l\leq L).
\end{equation}
Hence we obtain
\begin{equation}
  r_{n,-l}^{X,\mathcal{D}}
  =\frac{h_{\mathcal{D}\,n}}{h_{\mathcal{D}\,n-l}}r_{n-l,l}^{X,\mathcal{D}}
  \ \ (1\leq l\leq L).
  \label{r=h/hr}
\end{equation}
This result is obtained for an appropriate parameter range
(with which the inner product is well-defined) but the algebraic relations
\eqref{r=h/hr} themselves are valid for any parameter range.
If $r_{m,k}^{X,\mathcal{D}}$ ($0\leq m\leq n-1$, $-L\leq k\leq L$) are
known ($h_{\mathcal{D}\,n}$ are given in \eqref{hDn}), we obtain
$r_{n,k}^{X,\mathcal{D}}$ ($-L\leq k\leq -1$) by this relation (we can set
$r_{0,k}^{X,\mathcal{D}}=0$ ($-L\leq k\leq -1$)).
Therefore, in order to find the coefficients $r_{n,k}^{X,\mathcal{D}}$,
it is sufficient to find $r_{n,k}^{X,\mathcal{D}}$ ($n\geq 0$,
$0\leq k\leq L$).
The top coefficient $r_{n,L}^{X,\mathcal{D}}$ is easily obtained by comparing
the highest degree terms,
\begin{equation}
  r_{n,L}^{X,\mathcal{D}}
  =\frac{c^Xc^P_{\mathcal{D},n}}{c^P_{\mathcal{D},n+L}},
\end{equation}
where $c^X$ and $c^P_{\mathcal{D},n}$ are
\begin{align*}
  X(\eta)&=c^X\eta^L+(\text{lower order terms}),\\
  P_{\mathcal{D},n}(\eta)&=c^P_{\mathcal{D},n}\eta^{\ell_{\mathcal{D}}+n}
  +(\text{lower order terms}).
\end{align*}
The explicit forms of $c^P_{\mathcal{D},n}$ are found in \cite{rrmiop3} (L, J)
and \cite{os27} (W, AW).

\subsection{Generalized closure relations}
\label{sec:gcrmiop}

The recurrence relations with constant coefficients give rise to 
the generalized closure relations, for which we have the following:
\begin{conj}
For any polynomial $Y(\eta)$, we take $X(\eta)$ as \eqref{X}.
Then we have the closure relation of order $K=2L$ \eqref{crK} and
the eigenvalues of the matrix $A$ \eqref{matA,vecR} satisfy
\begin{equation}
  \alpha_1(z)>\alpha_2(z)>\cdots>\alpha_L(z)>0
  >\alpha_{L+1}(z)>\alpha_{L+2}(z)>\cdots>\alpha_{2L}(z)\ \ (z\geq 0).
  \label{alpha2}
\end{equation}
\label{conj_gcr}
\end{conj}
\vspace*{-8mm}
{\bf Remark 1}$\,$
{}From the form of the Hamiltonian \eqref{HoQM}--\eqref{HidQM},
the polynomials $R_i(z)=R^X_i(z)$ with coefficients $r_i^{(j)}=r_i^{X(j)}$
are
\begin{align}
  \text{L,\,J}:&\ \ R_i(z)=\sum_{j=0}^{[\frac12(K-i)]}r_i^{(j)}z^j
  \ \ (0\leq i\leq K-1),\quad
  R_{-1}(z)=\sum_{j=0}^{[\frac12K]}r_{-1}^{(j)}z^j,
  \label{RioQM}\\
  \text{W,\,AW}:&\ \ R_i(z)=\sum_{j=0}^{K-i}r_i^{(j)}z^j
  \ \ (0\leq i\leq K-1),\quad
  R_{-1}(z)=\sum_{j=0}^Kr_{-1}^{(j)}z^j,
  \label{RiidQM}
\end{align}
where $[a]$ denotes the greatest integer not exceeding $a$.

At present we do not have a proof of Conjecture\,\ref{conj_gcr} but we can
verify it for small values of $M$, $d_j$ and $\text{deg}\,Y(\eta)$
by direct calculation.
Such explicit examples will be given in the next section and
Appendix\,\ref{app:ex}.

In the rest of this section we assume that Conjecture\,\ref{conj_gcr} holds.
Then we have the exact Heisenberg operator solution of $X$ \eqref{X(t)} and
the creation/annihilation operators $a^{(j)}$ \eqref{a(j)}.
Action of \eqref{X(t)} on $\phi_{\mathcal{D}\,n}(x)$ is
\begin{equation*}
  e^{it\mathcal{H}}Xe^{-it\mathcal{H}}\phi_{\mathcal{D}\,n}(x)
  =\sum_{j=1}^{2L}e^{i\alpha_j(\mathcal{E}_n)t}a^{(j)}\phi_{\mathcal{D}\,n}(x)
  -R_{-1}(\mathcal{E}_n)R_0(\mathcal{E}_n)^{-1}\phi_{\mathcal{D}\,n}(x).
\end{equation*}
On the other hand the l.h.s. turns out to be
\begin{align*}
  e^{i\mathcal{H}t}Xe^{-i\mathcal{H}t}\phi_{\mathcal{D}\,n}(x)
  &=e^{i\mathcal{H}t}Xe^{-i\mathcal{E}_nt}\phi_{\mathcal{D}\,n}(x)
  =e^{-i\mathcal{E}_nt}e^{i\mathcal{H}t}
  \sum_{k=-L}^Lr_{n,k}^{X,\mathcal{D}}\phi_{\mathcal{D}\,n+k}(x)\n
  &=\sum_{k=-L}^Le^{i(\mathcal{E}_{n+k}-\mathcal{E}_n)t}\,
  r_{n,k}^{X,\mathcal{D}}\phi_{\mathcal{D}\,n+k}(x).
\end{align*}
Comparing these $t$-dependence (see \eqref{alpha2}) and using \eqref{Pinvj1},
we obtain
\begin{align}
  &\quad\ \ \alpha_{j}(\mathcal{E}_n)=\left\{
  \begin{array}{ll}
  \mathcal{E}_{n+L+1-j}-\mathcal{E}_n>0&(1\leq j\leq L)\\[2pt]
  \mathcal{E}_{n-(j-L)}-\mathcal{E}_n<0&(L+1\leq j\leq 2L)
  \end{array}\right.,
  \label{alphajEn}\\[2pt]
  &a^{(j)}\phi_{\mathcal{D}\,n}(x)=\left\{
  \begin{array}{ll}
  r_{n,L+1-j}^{X,\mathcal{D}}\phi_{\mathcal{D}\,n+L+1-j}(x)
  &(1\leq j\leq L)\\[6pt]
  r_{n,-(j-L)}^{X,\mathcal{D}}\phi_{\mathcal{D}\,n-(j-L)}(x)
  &(L+1\leq j\leq 2L)
  \end{array}\right.,
  \label{ajphiDn}\\[2pt]
  &-R_{-1}(\mathcal{E}_n)R_0(\mathcal{E}_n)^{-1}
  =r_{n,0}^{X,\mathcal{D}}.
  \label{Rm1En}
\end{align}
Therefore $a^{(j)}$ ($1\leq j\leq L$) and $a^{(j)}$ ($L+1\leq j\leq 2L$)
are creation and annihilation operators, respectively.
Among them $a^{(L)}$ and $a^{(L+1)}$ are fundamental,
$a^{(L)}\phi_{\mathcal{D},n}(x)\propto\phi_{\mathcal{D}\,n+1}(x)$ and
$a^{(L+1)}\phi_{\mathcal{D},n}(x)\propto\phi_{\mathcal{D}\,n-1}(x)$.

The creation and annihilation operators for eigenpolynomials are obtained
by the similarity transformation.
The similarity transformed Hamiltonian is
\begin{equation}
  \widetilde{\mathcal{H}}_{\mathcal{D}}\eqdef
  \Psi_{\mathcal{D}}(x)^{-1}\circ\mathcal{H}_{\mathcal{D}}
  \circ\Psi_{\mathcal{D}}(x),
  \ \ \widetilde{\mathcal{H}}_{\mathcal{D}}\check{P}_{\mathcal{D},n}(x)
  =\mathcal{E}_n\check{P}_{\mathcal{D},n}(x).
\end{equation}
Their explicit forms are
\begin{align}
  \text{L,\,J}:&\ \ \widetilde{\mathcal{H}}_{\mathcal{D}}(\bm{\lambda})
  =-4\biggl(c_2(\eta)\frac{d^2}{d\eta^2}
  +\Bigl(c_1(\eta,\bm{\lambda}^{[M_{\I},M_{\II}]})-2c_2(\eta)
  \frac{\partial_{\eta}\Xi_{\mathcal{D}}(\eta;\bm{\lambda})}
  {\Xi_{\mathcal{D}}(\eta;\bm{\lambda})}\Bigr)\frac{d}{d\eta}\n
  &\ \qquad\qquad\qquad
  +c_2(\eta)
  \frac{\partial^2_{\eta}\Xi_{\mathcal{D}}(\eta;\bm{\lambda})}
  {\Xi_{\mathcal{D}}(\eta;\bm{\lambda})}
  -c_1(\eta,\bm{\lambda}^{[M_{\I},M_{\II}]}-\bm{\delta})
  \frac{\partial_{\eta}\Xi_{\mathcal{D}}(\eta;\bm{\lambda})}
  {\Xi_{\mathcal{D}}(\eta;\bm{\lambda})}
  \biggr),
  \label{HtoQM}\\
  \text{W,\,AW}:&\ \ \widetilde{\mathcal{H}}_{\mathcal{D}}(\bm{\lambda})
  =V(x;\bm{\lambda}^{[M_{\I},M_{\II}]})\,
  \frac{\check{\Xi}_{\mathcal{D}}(x+i\frac{\gamma}{2};\bm{\lambda})}
  {\check{\Xi}_{\mathcal{D}}(x-i\frac{\gamma}{2};\bm{\lambda})}
  \biggl(e^{\gamma p}
  -\frac{\check{\Xi}_{\mathcal{D}}(x-i\gamma;\bm{\lambda}+\bm{\delta})}
  {\check{\Xi}_{\mathcal{D}}(x;\bm{\lambda}+\bm{\delta})}\biggr)\n
  &\ \qquad\qquad+V^*(x;\bm{\lambda}^{[M_{\I},M_{\II}]})\,
  \frac{\check{\Xi}_{\mathcal{D}}(x-i\frac{\gamma}{2};\bm{\lambda})}
  {\check{\Xi}_{\mathcal{D}}(x+i\frac{\gamma}{2};\bm{\lambda})}
  \biggl(e^{-\gamma p}
  -\frac{\check{\Xi}_{\mathcal{D}}(x+i\gamma;\bm{\lambda}+\bm{\delta})}
  {\check{\Xi}_{\mathcal{D}}(x;\bm{\lambda}+\bm{\delta})}\biggr),
  \label{HtidQM}
\end{align}
where we have written the parameter ($\bm{\lambda}$) dependence explicitly
(see \cite{rrmiop,rrmiop2,rrmiop3} for notation).
Corresponding to \eqref{crK}, this also satisfies the closure relation
\begin{equation}
  (\text{ad}\,\widetilde{\mathcal{H}})^KX
  =\sum_{i=0}^{K-1}(\text{ad}\,\widetilde{\mathcal{H}})^iX\cdot
  R_i(\widetilde{\mathcal{H}})
  +R_{-1}(\widetilde{\mathcal{H}}).
  \label{crKt}
\end{equation}
{}From the creation/annihilation operators
$a^{(j)}=a^{(j)}(\mathcal{H},X)$, we obtain the creation/annihilation
operators for eigenpolynomials,
\begin{align}
  &\tilde{a}^{(j)}\eqdef\Psi_{\mathcal{D}}(x)^{-1}\circ
  a^{(j)}(\mathcal{H},X)\circ\Psi_{\mathcal{D}}(x)
  =a^{(j)}(\widetilde{\mathcal{H}},X),\\
  &\tilde{a}^{(j)}\check{P}_{\mathcal{D},n}(x)=\left\{
  \begin{array}{ll}
  r_{n,L+1-j}^{X,\mathcal{D}}\check{P}_{\mathcal{D},n+L+1-j}(x)
  &(1\leq j\leq L)\\[6pt]
  r_{n,-(j-L)}^{X,\mathcal{D}}\check{P}_{\mathcal{D},n-(j-L)}(x)
  &(L+1\leq j\leq 2L)
  \end{array}\right..
  \label{ajtPDn}
\end{align}

\section{Examples}
\label{sec:ex}

In this section we present examples of the generalized closure relations
for the deformed systems described by the multi-indexed orthogonal
polynomials of Laguerre, Jacobi, Wilson and Askey-Wilson types.

We have verified Conjecture\,\ref{conj_gcr} for small $M$, $d_j$ and
$\text{deg}\,Y$ by direct calculation.
Such calculation suggests that the polynomials $R_i(z)$ ($0\leq i\leq K-1$)
depend on $L$ (or $K=2L$) only and the dependence of $d_j$ and $Y(\eta)$
enters in $R_{-1}(z)$.
We arrive at the following:
\begin{conj}
The eigenvalues of the matrix $A$ \eqref{matA,vecR} with $K=2L$ are given by
\eqref{alphaj_L}, \eqref{alphaj_J}, \eqref{alphaj_W} and \eqref{alphaj_AW}
for L, J, W and AW, respectively.
\label{conj_alphaj}
\end{conj}
{\bf Remark 1}$\,$
We can show that \eqref{alphaj_L}, \eqref{alphaj_J}, \eqref{alphaj_W} and
\eqref{alphaj_AW} satisfy \eqref{alpha2} and \eqref{alphajEn} for
appropriate parameter ranges.
These $\alpha_j(z)$'s depend on $d_j$ and $Y(\eta)$ only through the
degree of $X(\eta)$, $L=\ell_{\mathcal{D}}+\text{deg}\,Y(\eta)+1$.\\
{\bf Remark 2}$\,$
The polynomials $R_i(z)$ ($0\leq i\leq 2L-1$) are given by
\begin{equation}
  R_i(z)=(-1)^{i+1}\!\!\!\!\!\!\!\!\!\!\!
  \sum_{1\leq j_1<j_2<\cdots<j_{2L-i}\leq 2L}\!\!\!\!\!\!\!\!\!\!\!\!\!\!
  \alpha_{j_1}(z)\alpha_{j_2}(z)\cdots\alpha_{j_{2L-i}}(z)
  \ \ (0\leq i\leq 2L-1).
  \label{Riz}
\end{equation}
These expressions are symmetric under the exchange of $\alpha_j$ and
$\alpha_{2L+1-j}$.
For \eqref{alphaj_L}, \eqref{alphaj_J}, \eqref{alphaj_W} and
\eqref{alphaj_AW}, we can check that $\alpha_j+\alpha_{2L+1-j}$ and
$\alpha_j\alpha_{2L+1-j}$ are polynomials in $z$, see \eqref{al+al:J},
\eqref{al+al:W} and \eqref{al+al:AW}.
Hence $R_i(z)$ \eqref{Riz} are indeed polynomials in $z$.\\
{\bf Remark 3}$\,$
The above expressions \eqref{Riz} with \eqref{alphaj_L}, \eqref{alphaj_J},
\eqref{alphaj_W} and \eqref{alphaj_AW} are valid for $L=1$ case, namely the
original system ($\mathcal{D}=\{\}$, $\ell_{\mathcal{D}}=0$,
$\Xi_{\mathcal{D}}(\eta)=1$, $X(\eta)=X_{\text{min}}(\eta)=\eta$),
and the generalized closure relation reduces to the original closure relation.

If Conjecture\,\ref{conj_alphaj} holds, the unknown quantity of the
generalized closure relation is $R_{-1}(z)$ only.

\subsection{Multi-indexed Laguerre polynomials}
\label{sec:exL}

The data for the Laguerre polynomial are
\begin{equation}
  \mathcal{E}_n=4n,
  \ \ \tilde{\mathcal{E}}^{\I}_{\text{v}}=-4(g+\text{v}+\tfrac12),
  \ \ \tilde{\mathcal{E}}^{\II}_{\text{v}}=-4(g-\text{v}-\tfrac12),
  \ \ h_n=\tfrac{1}{2\,n!}\Gamma(n+g+\tfrac12).
  \label{dataL}
\end{equation}

The eigenvalues of the matrix $A$ \eqref{matA,vecR} with $K=2L$ are
conjectured as
\begin{equation}
  \alpha_j(z)=\left\{
  \begin{array}{ll}
  4(L+1-j)&(1\leq j\leq L)\\[2pt]
  -4(j-L)&(L+1\leq j\leq 2L)
  \end{array}\right.,
  \label{alphaj_L}
\end{equation}
which are constants.
It is trivial that these $\alpha_j$ satisfy \eqref{alpha2} and
\eqref{alphajEn}.
We have $R_i(z)=0$ ($i:\text{odd}>0$) due to $\alpha_j=-\alpha_{2L+1-j}$.
Note that $R_i(z)$ ($i:\text{even}\geq 0$) are also constants.

\medskip
\paragraph{Ex.1}
We explain $\mathcal{D}=\{1^{\I}\}$ (type $\I$) in detail as an illustration.
First we consider the lowest degree case $X(\eta)=X_{\text{min}}(\eta)$,
which corresponds to $Y(\eta)=1$.
The degree of $X(\eta)$ is $L=2$ and the data of the closure relation of
order $K=4$ are
\begin{align}
  X(\eta)&=X_{\text{min}}(\eta)=\tfrac12\eta(\eta+2g+1),\n
  R_0(z)&=-1024,\ \ R_1(z)=0,\ \ R_2(z)=80,\ \ R_3(z)=0,
  \label{L1R}\\
  R_{-1}(z)&=64\bigl(3z^2+2(10g+11)z+2(2g+1)(6g+13)\bigr),
  \nonumber
\end{align}
and $(\alpha_1,\alpha_2,\alpha_3,\alpha_4)=(8,4,-4,-8)$.
The creation operators, $a^{(1)}$ and $a^{(2)}$, and the annihilation
operators, $a^{(3)}$ and $a^{(4)}$, have the following forms,
\begin{align}
  a^{(1)}&=\tfrac{1}{768}\bigl(-128X-16(\text{ad}\,\mathcal{H})X
  +8(\text{ad}\,\mathcal{H})^2X+(\text{ad}\,\mathcal{H})^3X
  +\tfrac18R_{-1}(\mathcal{H})\bigr),\n
  a^{(2)}&=\tfrac{-1}{384}\bigl(-256X-64(\text{ad}\,\mathcal{H})X
  +4(\text{ad}\,\mathcal{H})^2X+(\text{ad}\,\mathcal{H})^3X
  +\tfrac14R_{-1}(\mathcal{H})\bigr),\n
  a^{(3)}&=\tfrac{1}{384}\bigl(\phantom{-}256X-64(\text{ad}\,\mathcal{H})X
  -4(\text{ad}\,\mathcal{H})^2X+(\text{ad}\,\mathcal{H})^3X
  -\tfrac14R_{-1}(\mathcal{H})\bigr),
  \label{L1a}\\
  a^{(4)}&=\tfrac{-1}{768}\bigl(\phantom{-}128X-16(\text{ad}\,\mathcal{H})X
  -8(\text{ad}\,\mathcal{H})^2X+(\text{ad}\,\mathcal{H})^3X
  -\tfrac18R_{-1}(\mathcal{H})\bigr).
  \nonumber
\end{align}
Here differential operators $\mathcal{H}$, $(\text{ad}\,\mathcal{H})X$ etc.
are
\begin{align}
  &\mathcal{H}_{\mathcal{D}}=-\frac{d^2}{dx^2}+U_{\mathcal{D}}(x),\quad
  X=X\bigl(\eta(x)\bigr)=\check{X}(x),\quad
  f'=\frac{df}{dx},\quad f^{(k)}=\frac{d^kf}{dx^k}\n
  &(\text{ad}\,\mathcal{H})X=-2\check{X}'\frac{d}{dx}-\check{X}'',\quad
  (\text{ad}\,\mathcal{H})^2X=4\check{X}''\frac{d^2}{dx^2}
  +4\check{X}'''\frac{d}{dx}+\check{X}^{(4)}+2\check{X}'U'_{\mathcal{D}}(x),\n
  &(\text{ad}\,\mathcal{H})^3X=-8\check{X}'''\frac{d^3}{dx^3}
  -12\check{X}^{(4)}\frac{d^2}{dx^2}
  -2\bigl(3\check{X}^{(5)}+6\check{X}''U_{\mathcal{D}}'(x)
  +2\check{X}'U_{\mathcal{D}}''(x)\bigr)\frac{d}{dx}\n
  &\phantom{(\text{ad}\,\mathcal{H})^3X=}
  -\check{X}^{(6)}-6\check{X}'''U_{\mathcal{D}}'(x)
  -8\check{X}''U_{\mathcal{D}}''(x)-2\check{X}'U_{\mathcal{D}}'''(x),
\end{align}
and the potential $U_{\mathcal{D}}(x)$ in this case is
\begin{equation}
  U_{\mathcal{D}}(x)=x^2+\frac{g(g+1)}{x^2}-2g-3+\frac{4}{x^2+g+\frac12}
  -\frac{4(2g+1)}{(x^2+g+\frac12)^2}.
\end{equation}
The eigenfunctions of $\mathcal{H}_{\mathcal{D}}$ are
\begin{align}
  \phi_{\mathcal{D}\,n}(x)
  &=\frac{2e^{-\frac12x^2}x^{g+1}}{x^2+g+\frac12}P_{\mathcal{D},n}(x^2),\n
  P_{\mathcal{D},n}(\eta)
  &=(g+\tfrac12+\eta)\partial_{\eta}L^{(g-\frac12)}_n(\eta)
  -(g+\tfrac32+\eta)L^{(g-\frac12)}_n(\eta).
\end{align}
The coefficients of the 5-term recurrence relations for
$X(\eta)=X_{\text{min}}(\eta)$ are \cite{d14,mt14,rrmiop2}
\begin{align}
  r_{n,2}^{X,\mathcal{D}}&=\tfrac12(n+1)(n+2),\quad
  r_{n,-2}^{X,\mathcal{D}}=\tfrac18(2g+2n-3)(2g+2n+3),\n
  r_{n,1}^{X,\mathcal{D}}&=-(n+1)(2g+2n+3),\quad
  r_{n,-1}^{X,\mathcal{D}}=-\tfrac12(2g+2n-1)(2g+2n+3),
  \label{Ex1_L}\\
  r_{n,0}^{X,\mathcal{D}}
  &=\tfrac18\bigl(24n^2+4(10g+11)n+(2g+1)(6g+13)\bigr).
  \nonumber
\end{align}
We can check \eqref{alphajEn}--\eqref{Rm1En}, \eqref{ajtPDn}, \eqref{r=h/hr},
etc.

Next we take $Y(\eta)=\eta$. Then we have $L=3$ and the data of the closure
relation of order $K=6$ are
\begin{align}
  X(\eta)&=\eta^2\bigl(\tfrac13\eta+\tfrac14(2g+1)\bigr),\n
  R_0(z)&=147456,\ \ R_2(z)=-12544,\ \ R_4(z)=224,
  \ \ R_1(z)=R_3(z)=R_5(z)=0,
  \label{L1yR}\\
  R_{-1}(z)&=-1536\bigl(10z^3+3(26g+33)z^2+2(84g^2+240g+139)z\n
  &\quad+2(2g+1)(2g+5)(10g+27)\bigr),
  \nonumber
\end{align}
and $(\alpha_1,\alpha_2,\alpha_3,\alpha_4,\alpha_5,\alpha_6)
=(12,8,4,-4,-8,-12)$.

For $Y(\eta)=\eta^2$, we have $L=4$ and the data of the closure
relation of order $K=8$ are
\begin{align}
  X(\eta)&=\eta^3\bigl(\tfrac14\eta+\tfrac16(2g+1)\bigr),\n
  R_0(z)&=-37748736,\ \ R_2(z)=3358720,\ \ R_4(z)=-69888,\ \ R_6(z)=480,\n
  R_1(z)&=R_3(z)=R_5(z)=R_7(z)=0,
  \label{L1y2R}\\
  R_{-1}(z)&=24576\bigl(105z^4+20(50g+67)z^3+60(52g^2+152g+107)z^2\n
  &\quad+32(6g+5)(18g^2+77g+94)z+8(2g+1)(2g+5)(2g+7)(14g+45)\bigr),
  \nonumber
\end{align}
and $(\alpha_1,\alpha_2,\ldots,\alpha_8)
=(16,12,8,4,-4,-8,-12-16)$.

\medskip
\paragraph{Ex.2}
For $\mathcal{D}=\{1^{\II}\}$ (type $\II$) and $Y(\eta)=1$,
we have $L=2$ and the data of the closure relation of order $K=4$ are
\begin{align}
  X(\eta)&=X_{\text{min}}(\eta)=-\tfrac12\eta(\eta+2g-3),\n
  R_0(z)&=-1024,\ \ R_1(z)=0,\ \ R_2(z)=80,\ \ R_3(z)=0,
  \label{L1IIR}\\
  R_{-1}(z)&=-64\bigl(3z^2+2(10g-9)z+2(2g-3)(6g+1)\bigr),
  \nonumber
\end{align}
and $(\alpha_1,\alpha_2,\alpha_3,\alpha_4)=(8,4,-4,-8)$.

More examples are presented in Appendix \ref{app:ex}.

\subsection{Multi-indexed Jacobi polynomials}
\label{sec:exJ}

The data for the Jacobi polynomial are
\begin{align}
  &\mathcal{E}_n=4n(n+g+h),
  \ \ h_n=\frac{\Gamma(n+g+\frac12)\Gamma(n+h+\frac12)}
  {2\,n!\,(2n+g+h)\Gamma(n+g+h)},\n
  &\tilde{\mathcal{E}}^{\I}_{\text{v}}
  =-4(g+\text{v}+\tfrac12)(h-\text{v}-\tfrac12),
  \ \ \tilde{\mathcal{E}}^{\II}_{\text{v}}
  =-4(g-\text{v}-\tfrac12)(h+\text{v}+\tfrac12),
  \label{dataJ}
\end{align}
and we set $a=g+h$ and $b=g-h$.

The eigenvalues of the matrix $A$ \eqref{matA,vecR} with $K=2L$ are
conjectured as
\begin{equation}
  \alpha_j(z)=\left\{
  \begin{array}{ll}
  4(L+1-j)^2+4(L+1-j)\sqrt{z+a^2}&(1\leq j\leq L)\\[4pt]
  4(j-L)^2-4(j-L)\sqrt{z+a^2}&(L+1\leq j\leq 2L)
  \end{array}\right..
  \label{alphaj_J}
\end{equation}
Remark that $\alpha_j(\mathcal{E}_n)$ is square root free,
$\sqrt{\mathcal{E}_n+a^2}=2n+a$.
We can show that these $\alpha_j$ satisfy \eqref{alpha2} and
\eqref{alphajEn} for $a>2L-1$.
Note that
\begin{align}
  \alpha_j(z)+\alpha_{2L+1-j}(z)&=\left\{
  \begin{array}{ll}
  8(L+1-j)^2&(1\leq j\leq L)\\[2pt]
  8(j-L)^2&(L+1\leq j\leq 2L)
  \end{array}\right.,
  \label{al+al:J}\\
  \alpha_j(z)\alpha_{2L+1-j}(z)&=\left\{
  \begin{array}{ll}
  16(L+1-j)^2\bigl((L+1-j)^2-z-a^2\bigr)&(1\leq j\leq L)\\[2pt]
  16(j-L)^2\bigl((j-L)^2-z-a^2\bigr)&(L+1\leq j\leq 2L)
  \end{array}\right..\nonumber
\end{align}

\medskip
\paragraph{Ex.1}
First we consider $\mathcal{D}=\{1^{\I}\}$ (type $\I$) and take
$X=X_{\text{min}}$ ($\Rightarrow\ell_{\mathcal{D}}=1$, $L=2$, $K=4$),
\begin{equation}
  X(\eta)=X_{\text{min}}(\eta)=
  \tfrac14\eta\bigl((b+2)\eta+2(a-1)\bigr).
\end{equation}
The closure relation of order 4 holds with the following $R_i$:
\begin{align}
  R_0(z)&=-1024(z+a^2-1)(z+a^2-4),
  \ \ R_1(z)=-1024(z+a^2-\tfrac52),\n
  R_2(z)&=80(z+a^2-\tfrac{33}{5}),\ \ R_3(z)=40,\n
  R_{-1}(z)&=128(b+2)\Bigl(z^2-\bigl((b+2)^2+3a^2-10a+1\bigr)z
  \label{J1R}\\
  &\qquad\qquad\qquad
  +2(a-1)(a-2)\bigl((b+2)^2-2a^2-a-3)\bigr)\Bigr).
  \nonumber
\end{align}
The eigenvalues of $A$ \eqref{matA,vecR} are
\begin{align*}
  &\alpha_1(\mathcal{H})=16+8\sqrt{\mathcal{H}+a^2},
  \ \ \alpha_2(\mathcal{H})=4+4\sqrt{\mathcal{H}+a^2},\\
  &\alpha_4(\mathcal{H})=16-8\sqrt{\mathcal{H}+a^2},
  \ \ \alpha_3(\mathcal{H})=4-4\sqrt{\mathcal{H}+a^2},
\end{align*}
and $\alpha_j(\mathcal{E}_n)$ is square root free,
$\sqrt{\mathcal{E}_n+a^2}=2n+a$.
We can check \eqref{alpha} for $a>3$.
Therefore $a^{(1)}$ and $a^{(2)}$ are the creation operators, and
$a^{(3)}$ and $a^{(4)}$ are the annihilation operators.
The potential and eigenfunctions are
\begin{align}
  U_{\mathcal{D}}(x)&=\frac{g(g+1)}{\sin^2x}+\frac{(h-1)(h-2)}{\cos^2x}
  -a^2+\frac{8(a-1)}{a-1+(b+2)\cos 2x}
  -\frac{8(2g+1)(2h-3)}{\bigl(a-1+(b+2)\cos 2x\bigr)^2},\n
  \phi_{\mathcal{D}\,n}(x)
  &=\frac{-8(\sin x)^{g+1}(\cos x)^{h-1}}{a-1+(b+2)\cos 2x}
  P_{\mathcal{D},n}(\cos 2x),\n
  P_{\mathcal{D},n}(\eta)
  &=\tfrac14(1+\eta)\bigl(a-1+(b+2)\eta\bigr)
  \partial_{\eta}P^{(g-\frac12,h-\frac12)}_n(\eta)\n
  &\quad-\tfrac14(\tfrac32-h)\bigl(a+1+(b+2)\eta\bigr)
  P^{(g-\frac12,h-\frac12)}_n(\eta).
\end{align}
The coefficients of the 5-term recurrence relations are \cite{rrmiop2}
\begin{align}
  r_{n,2}^{X,\mathcal{D}}&=
  \frac{(n+1)_2(b+2)(a+n)_2(2h+2n-3)}{(a+2n)_4(2h+2n+1)},\n
  r_{n,-2}^{X,\mathcal{D}}&=
  \frac{(b+2)(2g+2n-3)(2g+2n+3)(h+n-\tfrac32)_2}{4(a+2n-3)_4},\n
  r_{n,1}^{X,\mathcal{D}}&=
  \frac{(n+1)(a-1)(a+n)(2g+2n+3)(2h+2n-3)}{(a+2n-1)_3(a+2n+3)},
  \label{Ex1_J}\\
  r_{n,-1}^{X,\mathcal{D}}&=
  \frac{(a-1)(2g+2n-1)(2g+2n+3)(h+n-\tfrac32)_2}{(a+2n-3)(a+2n-1)_3},\n
  r_{n,0}^{X,\mathcal{D}}&=
  \frac{b+2}{4(a+2n-2)_2(a+2n+1)_2}\Bigl(
  -b(b+4)\bigl(2n(a+n)-(a-2)(a-1)\bigr)\n
  &\qquad\qquad
  +(a+2n-1)(a+2n+1)\bigl(2n(a+n)-(a-2)(2a-1)\bigr)\Bigr).
  \nonumber
\end{align}
We can check \eqref{alphajEn}--\eqref{Rm1En}, \eqref{ajtPDn}, \eqref{r=h/hr},
etc.

\paragraph{Ex.2}
Next we consider $\mathcal{D}=\{1^{\II}\}$ (type $\II$) and take
$X=X_{\text{min}}$.
Since these two cases $\mathcal{D}=\{1^{\I}\}$ and $\mathcal{D}=\{1^{\II}\}$
are essentially same \cite{os16}, we present $R_{-1}(z)$ only,
\begin{align}
  R_{-1}(z)&=-128(b-2)\Bigl(z^2-\bigl((b-2)^2+3a^2-10a+1\bigr)z\n
  &\qquad\qquad\qquad
  +2(a-1)(a-2)\bigl((b-2)^2-2a^2-a-3)\bigr)\Bigr).
\end{align}

More examples are presented in Appendix \ref{app:ex}.

\subsection{Multi-indexed Wilson polynomials}
\label{sec:exW}

The data for the Wilson polynomial are
\begin{align}
  &\mathcal{E}_n=n(n+b_1-1),\ \ b_1\eqdef\sum_{i=1}^4a_i,
  \ \ h_n=\frac{2\pi n!\,(n+b_1-1)_n
  \prod_{1\leq i<j\leq 4}\Gamma(n+a_i+a_j)}
  {\Gamma(2n+b_1)},\n
  &\tilde{\mathcal{E}}^{\I}_{\text{v}}
  =-(a_1+a_2-\text{v}-1)(a_3+a_4+\text{v}),
  \ \ \tilde{\mathcal{E}}^{\II}_{\text{v}}
  =-(a_3+a_4-\text{v}-1)(a_1+a_2+\text{v}),
  \label{dataW}
\end{align}
and we set $\sigma_1=a_1+a_2$, $\sigma_2=a_1a_2$, $\sigma'_1=a_3+a_4$,
$\sigma'_2=a_3a_4$, $b_2=\!\!\!\!\sum\limits_{1\leq i<j\leq 4}\!\!\!a_ia_j$,
$b_3=\!\!\!\!\sum\limits_{1\leq i<j<k\leq 4}\!\!\!a_ia_ja_k$ and
$b_4=a_1a_2a_3a_4$.

The eigenvalues of the matrix $A$ \eqref{matA,vecR} with $K=2L$ are
conjectured as
\begin{equation}
  \alpha_j(z)=\left\{
  \begin{array}{ll}
  (L+1-j)^2+(L+1-j)\sqrt{4z+(b_1-1)^2}&(1\leq j\leq L)\\[4pt]
  (j-L)^2-(j-L)\sqrt{4z+(b_1-1)^2}&(L+1\leq j\leq 2L)
  \end{array}\right..
  \label{alphaj_W}
\end{equation}
Remark that $\alpha_j(\mathcal{E}_n)$ is square root free,
$\sqrt{4\mathcal{E}_n+(b_1-1)^2}=2n+b_1-1$.
We can show that these $\alpha_j$ satisfy \eqref{alpha2} and
\eqref{alphajEn} for $b_1>2L$.
Note that
\begin{align}
  \alpha_j(z)+\alpha_{2L+1-j}(z)&=\left\{
  \begin{array}{ll}
  2(L+1-j)^2&(1\leq j\leq L)\\[2pt]
  2(j-L)^2&(L+1\leq j\leq 2L)
  \end{array}\right.,
  \label{al+al:W}\\
  \alpha_j(z)\alpha_{2L+1-j}(z)&=\left\{
  \begin{array}{ll}
  (L+1-j)^2\bigl((L+1-j)^2-4z-(b_1-1)^2\bigr)&(1\leq j\leq L)\\[2pt]
  (j-L)^2\bigl((j-L)^2-4z-(b_1-1)^2\bigr)&(L+1\leq j\leq 2L)
  \end{array}\right..\nonumber
\end{align}

\medskip
\paragraph{Ex.1}
We consider $\mathcal{D}=\{1^{\I}\}$ and take $X=X_{\text{min}}$
($\Rightarrow\ell_{\mathcal{D}}=1$, $L=2$, $K=4$),
\begin{equation}
  X(\eta)=X_{\text{min}}(\eta)=\tfrac14\eta\bigl(2(\sigma_1-\sigma'_1-2)\eta
  +4(\sigma_2\sigma'_1-\sigma_1\sigma'_2-\sigma_1\sigma'_1+2\sigma'_2)
  +\sigma_1+3\sigma'_1-2\bigr).
\end{equation}
The closure relation of order 4 holds with the following $R_i$: $R_i(z)$
($0\leq i\leq 4$) are given by Conjecture \ref{conj_alphaj} and \eqref{Riz},
\begin{align}
  R_3(z)&=10,\quad
  R_2(z)=5z'-33,\quad z'\eqdef 4z+(b_1-1)^2,\n
  R_1(z)&=-8(2z'-5),\quad R_0(z)=-4(z'-1)(z'-4),
  \label{W1R}
\end{align}
and $R_{-1}(z)$ is presented in \eqref{W:1I} because of its lengthy expression.
The coefficients of the 5-term recurrence relations are found in
\cite{rrmiop2} and we can check \eqref{alphajEn}--\eqref{Rm1En},
\eqref{ajtPDn}, \eqref{r=h/hr}, etc.

\subsection{Multi-indexed Askey-Wilson polynomials}
\label{sec:exAW}

The data for the Askey-Wilson polynomial are
\begin{align}
  &\mathcal{E}_n=(q^{-n}-1)(1-b_4q^{n-1}),\ \ b_4\eqdef\prod_{i=1}^4a_i,
  \ \ h_n=\frac{2\pi(b_4q^{n-1};q)_n(b_4q^{2n};q)_{\infty}}
  {(q^{n+1};q)_{\infty}\prod_{1\leq i<j\leq 4}(a_ia_jq^n;q)_{\infty}},\n
  &\tilde{\mathcal{E}}^{\I}_{\text{v}}
  =-(1-a_1a_2q^{-\text{v}-1})(1-a_3a_4q^{\text{v}}),
  \ \ \tilde{\mathcal{E}}^{\II}_{\text{v}}
  =-(1-a_3a_4q^{-\text{v}-1})(1-a_1a_2q^{\text{v}}),
  \label{dataAW}
\end{align}
and we set $\sigma_1=a_1+a_2$, $\sigma_2=a_1a_2$, $\sigma'_1=a_3+a_4$,
$\sigma'_2=a_3a_4$, $b_1=\sum\limits_{i=1}^4a_i$,
$b_2=\!\!\!\!\sum\limits_{1\leq i<j\leq 4}\!\!\!a_ia_j$ and
$b_3=\!\!\!\!\sum\limits_{1\leq i<j<k\leq 4}\!\!\!a_ia_ja_k$.

The eigenvalues of the matrix $A$ \eqref{matA,vecR} with $K=2L$ are
conjectured as
\begin{equation}
  \alpha_j(z)=\left\{
  \begin{array}{ll}
  \tfrac12\bigl((q^{-\frac12(L+1-j)}-q^{\frac12(L+1-j)})^2
  (z+1+q^{-1}b_4)\\[4pt]
  \quad
  +(q^{-(L+1-j)}-q^{L+1-j})\sqrt{(z+1+q^{-1}b_4)^2-4q^{-1}b_4}\,\bigr)
  &(1\leq j\leq L)\\[4pt]
  \tfrac12\bigl((q^{-\frac12(j-L)}-q^{\frac12(j-L)})^2
  (z+1+q^{-1}b_4)\\[4pt]
  \quad
  -(q^{-(j-L)}-q^{j-L})\sqrt{(z+1+q^{-1}b_4)^2-4q^{-1}b_4}\,\bigr)
  &\!\!\!\!\!\!\!(L+1\leq j\leq 2L)
  \end{array}\right.\!\!\!.\!\!
  \label{alphaj_AW}
\end{equation}
Remark that $\alpha_j(\mathcal{E}_n)$ is square root free,
$\sqrt{(\mathcal{E}_n+1+q^{-1}b_4)^2-4q^{-1}b_4}=q^{-n}-b_4q^{n-1}$.
We can show that these $\alpha_j$ satisfy \eqref{alpha2} and
\eqref{alphajEn} for $b_4<q^{2L}$.
Note that
\begin{align}
  &\alpha_j(z)+\alpha_{2L+1-j}(z)=\left\{
  \begin{array}{ll}
  (q^{-\frac12(L+1-j)}-q^{\frac12(L+1-j)})^2(z+1+q^{-1}b_4)
  &(1\leq j\leq L)\\[2pt]
  (q^{-\frac12(j-L)}-q^{\frac12(j-L)})^2(z+1+q^{-1}b_4)
  &(L+1\leq j\leq 2L)
  \end{array}\right.,\n
  &\alpha_j(z)\alpha_{2L+1-j}(z)
 \label{al+al:AW}\\
  &=\left\{
  \begin{array}{ll}
  (q^{-\frac12(L+1-j)}-q^{\frac12(L+1-j)})^2\\[2pt]
  \quad\times
  \bigl((q^{-\frac12(L+1-j)}+q^{\frac12(L+1-j)})^2q^{-1}b_4
  -(z+1+q^{-1}b_4)^2\bigr)
  &(1\leq j\leq L)\\[4pt]
  (q^{-\frac12(j-L)}-q^{\frac12(j-L)})^2\\[2pt]
  \quad\times
  \bigl((q^{-\frac12(j-L)}+q^{\frac12(j-L)})^2q^{-1}b_4
  -(z+1+q^{-1}b_4)^2\bigr)
  &(L+1\leq j\leq 2L)
  \end{array}\right..\nonumber
\end{align}

\medskip
\paragraph{Ex.1}
We consider $\mathcal{D}=\{1^{\I}\}$ and take $X=X_{\text{min}}$
($\Rightarrow\ell_{\mathcal{D}}=1$, $L=2$, $K=4$),
\begin{equation}
  X(\eta)=X_{\text{min}}(\eta)=\frac{\eta}{(1+q)\sigma_2}
  \Bigl(2q^{\frac12}(\sigma_2-\sigma'_2q^2)\eta
  -(1+q)\bigl(\sigma_1(1-\sigma'_2)q+\sigma'_1(\sigma_2-q^2)\bigr)\Bigr).
\end{equation}
The closure relation of order 4 holds with the following $R_i$: $R_i(z)$
($0\leq i\leq 4$) are given by Conjecture \ref{conj_alphaj} and \eqref{Riz},
\begin{align}
  R_3(z)&=q^{-2}(1-q)^2(1+3q+q^2)z',\quad z'\eqdef z+1+q^{-1}b_4,\n
  R_2(z)&=-q^{-3}(1-q)^2\bigl(
  (1-q-5q^2-q^3+q^4)z^{\prime\,2}+q^{-2}(1+q)^2(1+3q^2+q^4)b_4\bigr),\n
  R_1(z)&=-q^{-3}(1-q)^4(1+q)^2z'\bigl(2z^{\prime\,2}
  -q^{-3}(1+q+4q^2+q^3+q^4)b_4\bigr),
  \label{AW1R}\\
  R_0(z)&=-q^{-3}(1-q)^4(1+q)^2\bigl(z^{\prime\,2}-q^{-2}(1+q)^2b_4\bigr)
  \bigl(z^{\prime\,2}-q^{-3}(1+q^2)^2b_4\bigr),
  \nonumber
\end{align}
and $R_{-1}(z)$ is presented in \eqref{AW:1I} because of its lengthy expression.
The coefficients of the 5-term recurrence relations are found in
\cite{rrmiop2} and we can check \eqref{alphajEn}--\eqref{Rm1En},
\eqref{ajtPDn}, \eqref{r=h/hr}, etc.

\section{Summary and Comments}
\label{sec:summary}

Exactly solvable quantum mechanical systems described by Laguerre, Jacobi,
Wilson and Askey-Wilson polynomials satisfy the shape invariance and closure
relation. The shape invariance is inherited by the deformed systems described
by the multi-indexed orthogonal polynomials of Laguerre, Jacobi, Wilson and
Askey-Wilson types. The closure relation is not inherited, because it is
related to the three term recurrence relations for the ordinary orthogonal
polynomials. These multi-indexed orthogonal polynomials satisfy the $1+2L$
($L\geq 2$) term recurrence relations with constant coefficients \eqref{XP},
which is obtained from multiplication by $X$.
We generalize the closure relation \eqref{crK} and discuss that the deformed
systems satisfy the generalized closure relations, which corresponds to the
$1+2L$ term recurrence relations with constant coefficients.
This is stated as Conjecture \ref{conj_gcr}.
The generalized closure relation gives the exact Heisenberg operator solution
of $X$, and its negative/positive frequency parts provide the
creation/annihilation operators of the system.
For small $L$ cases, we can verify the generalized closure relations.
Some examples and Conjecture \ref{conj_alphaj} are presented.
Although we have discussed Laguerre, Jacobi, Wilson and Askey-Wilson cases,
the method is applicable to other case (1) and case (2)
exceptional/multi-indexed polynomials.
For example, the exceptional Hermite polynomials with multi-indices are
worth for investigation.

In contrast to the shape invariance which is applicable to quantum system only,
the (generalized) closure relation is applicable to not only quantum system
but also classical system \cite{os7}.
The commutator $[\cdot\,,\cdot]$ in quantum mechanics becomes the Poisson
bracket $\{\cdot\,,\cdot\}_{\text{PB}}$ in classical mechanics.
To obtain the classical limit, we have to recover $\hbar$ (reduced Planck
constant) dependence, because we have used the $\hbar=1$ unit.
Then the commutator $\frac{1}{i\hbar}[\cdot\,,\cdot]$ and the Hamiltonian
$\mathcal{H}$ become the Poisson bracket $\{\cdot\,,\cdot\}_{\text{PB}}$
and the classical Hamiltonian $\mathcal{H}^{\text{cl}}$ respectively,
in the $\hbar\to 0$ limit.
The time evolution equation of a phase space function $F$ is
$\frac{d}{dt}F=-\{\mathcal{H}^{\text{cl}},F\}_{\text{PB}}$ and it is
solved as $F(t)=\sum\limits_{n=0}^{\infty}\frac{(-t)^n}{n!}
(\text{ad}_{\text{PB}}\mathcal{H}^{\text{cl}})^nF$, where
$\text{ad}_{\text{PB}}\mathcal{H}^{\text{cl}}$ stands for the operation
$(\text{ad}_{\text{PB}}\mathcal{H}^{\text{cl}})F=
\{\mathcal{H}^{\text{cl}},F\}_{\text{PB}}$.
By the similar calculation in \S\,\ref{sec:gcr}, the generalized closure
relation gives us the time evolution of $X$, $X(t)$, explicitly.

For the ordinary orthogonal polynomials, any recurrence relations of the
form $X(\eta)P_n(\eta)=\sum\limits_{k=-L}^Lr^X_{n,k}P_{n+k}(\eta)$ can be
obtained from the three term recurrence relations. On the other hand,
for multi-indexed orthogonal polynomials of L, J, W and AW types, different
choices of $X$ give different recurrence relations with constant coefficients
\eqref{XP} in general. Correspondingly we have infinitely many
creation/annihilation operators. It is a challenging problem to study
their relations. Even for the simplest choice $X=X_{\text{min}}$,
there are $L$ creation operators $a^{(j)}$ ($1\leq j\leq L$) and $L$
annihilation operators $a^{(j)}$ ($L+1\leq j\leq 2L$) and it is a good
problem to study their relations (commutators etc.).
In harmonic oscillator, the coherent state is obtained as an eigenstate
of the annihilation operator. It is also a challenging problem to find
the eigenstates of the annihilation operators $a^{(j)}$ ($L+1\leq j\leq 2L$).

\section*{Acknowledgments}

I thank R.\,Sasaki for discussion and reading of the manuscript.
I am supported in part by Grant-in-Aid for Scientific Research
from the Ministry of Education, Culture, Sports, Science and Technology
(MEXT), No.25400395.

\bigskip
\appendix
\section{Diagonalization of Some Matrix}
\label{app:diag}

In this appendix we present some formulas for the diagonalization of a matrix
\eqref{matA}.

Let us consider a $K\times K$ matrix $A$,
\begin{equation}
  A=\begin{pmatrix}
  0&&&&&R_0\\
  1&0&&\text{\Large$O$}&&R_1\\
  &1&0&&&R_2\\
  &&\ddots&\ddots&&\vdots\\
  &\text{\Large$O$}&&1&0&R_{K-2}\\
  &&&&1&R_{K-1}
  \end{pmatrix},
  \label{matA}
\end{equation}
where $R_i$'s are real numbers.
Its characteristic polynomial is
\begin{equation}
   |xE-A|=x^K-\sum_{i=0}^{K-1}R_ix^i.
\end{equation}
We assume that the matrix $A$ has $K$ distinct real non-vanishing eigenvalues,
\begin{equation}
  |xE-A|=\prod_{j=1}^K(x-\alpha_j),\quad
  \alpha_1>\alpha_2>\cdots>\alpha_K,\quad\alpha_i\neq 0.
\end{equation}
Then $R_i$ is expressed by $\alpha_j$,
\begin{equation}
  R_i=(-1)^{K-i-1}\!\!\!\!\!\!\!\!\!\!\!
  \sum_{1\leq j_1<j_2<\cdots<j_{K-i}\leq K}\!\!\!\!\!\!\!\!\!\!\!\!\!\!
  \alpha_{j_1}\alpha_{j_2}\cdots\alpha_{j_{K-i}}
  \ \ (0\leq i\leq K-1).
\end{equation}
The eigenvector $\vec{p}_j=(p_{ij})_{1\leq i\leq K}$ corresponding to
the eigenvalue $\alpha_j$ is
\begin{equation}
  p_{ij}=\alpha_j^{K-i}-\sum_{k=1}^{K-i}R_{K-k}\,\alpha_j^{K-i-k}.
\end{equation}
(Here our summation convention is $\sum_{i=n}^{n-1}*=0$.)
Note that $p_{Kj}=1$.
The matrix $A$ is diagonalized as
\begin{equation}
  P^{-1}AP=\text{diag}(\alpha_1,\alpha_2,\ldots,\alpha_K),\quad
  P=(p_{ij})_{1\leq i,j\leq K}
  =\bigl(\vec{p}_1\ \vec{p}_2\ \cdots\ \vec{p}_K\bigr).
\end{equation}
The determinant of $P$ is
\begin{equation}
  |P|=\prod_{1\leq i<j\leq K}(\alpha_i-\alpha_j)\neq 0,
\end{equation}
and the matrix elements of $P^{-1}$ are
\begin{equation}
  (P^{-1})_{ji}=\alpha_j^{i-1}
  \prod_{\genfrac{}{}{0pt}{}{k=1}{k\neq j}}^K(\alpha_j-\alpha_k)^{-1}.
\end{equation}
Note that
\begin{equation}
  \sum_{j=1}^K\alpha_j^{-1}(P^{-1})_{j1}
  =\sum_{j=1}^K
  \frac{1}{\alpha_j\prod\limits_{\genfrac{}{}{0pt}{}{k=1}{k\neq j}}^K
  (\alpha_j-\alpha_k)}
  =\frac{(-1)^{K-1}}{\prod\limits_{j=1}^K\alpha_j}
  =R_0^{-1}.
\end{equation}

\section{More Examples}
\label{app:ex}

In this appendix we present more examples of the generalized closure
relations, which support Conjecture \ref{conj_gcr} and Conjecture
\ref{conj_alphaj}.
We take $X(\eta)=X_{\text{min}}(\eta)$ ($Y(\eta)=1$). The order of the
generalized closure relation is $K=2L=2(\ell_{\mathcal{D}}+1)$.
Since $R_i(z)$ ($0\leq i\leq 2L-1$) are obtained by Conjecture
\ref{conj_alphaj} and \eqref{Riz}, we present $R_{-1}(z)$ only.

\subsection{Laguerre type}
\label{app:L}

We present $R_{-1}(z)$ for all $\mathcal{D}$ with $K\leq 8$.

\noindent
\underline{$K=2$}
\begin{equation}
  \mathcal{D}=\{\}:
  \ \ R_{-1}(z)=-8(z+2g+1).
\end{equation}
\underline{$K=4$}
\begin{align}
  \mathcal{D}&=\{1^{\I}\}:
  \ \ R_{-1}(z)=64\bigl(2(1+2g)(13+6g)+2(11+10g)z+3z^2\bigr),\\
  \mathcal{D}&=\{1^{\II}\}:
  \ \ R_{-1}(z)=-64\bigl(2(2g-3)(1+6g)+2(10g-9)z+3z^2\bigr).
\end{align}
\underline{$K=6$}
\begin{align}
  \mathcal{D}&=\{2^{\I}\}:
  \ \ R_{-1}(z)=-1536\bigl(2(1+2g)(3+2g)(41+14g)+4(61+108g+36g^2)z\n
  &\qquad\qquad\qquad\qquad\qquad\quad
  +24(3+2g)z^2+5z^3\bigr),\\
  \mathcal{D}&=\{2^{\II}\}:
  \ \ R_{-1}(z)=-1536\bigl(2(2g-5)(2g-1)(3+14g)+4(25-84g+36g^2)z\n
  &\qquad\qquad\qquad\qquad\qquad\quad
  +12(4g-5)z^2+5z^3\bigr),\\
  \mathcal{D}&=\{1^{\I},2^{\I}\}:
  \ \ R_{-1}(z)=-1536\bigl(2(1+2g)(5+2g)(45+14g)+4(97+132g+36g^2)z\n
  &\qquad\qquad\qquad\qquad\qquad\quad
  +12(7+4g)z^2+5z^3\bigr),\\
  \mathcal{D}&=\{1^{\II},2^{\II}\}:
  \ \ R_{-1}(z)=1536\bigl(2(2g-5)(2g-3)(14g-1)+4(61-108g+36g^2)z\n
  &\qquad\qquad\qquad\qquad\qquad\quad
  +24(2g-3)z^2+5z^3\bigr).
\end{align}
\underline{$K=8$}
\begin{align}
  \mathcal{D}&=\{3^{\I}\}:\n
  &R_{-1}(z)=12288(8(1+2g)(3+2g)(5+2g)(113+30g)\n
  &\qquad\qquad
  +16(935+2072g+1284g^2+224g^3)z+4(1405+1848g+492g^2)z^2\n
  &\qquad\qquad
  +20(41+22g)z^3+35z^4\bigr),\\
  \mathcal{D}&=\{3^{\II}\}:\n
  &R_{-1}(z)=-12288\bigl(24(2g-7)(2g-1)(-5-16g+20g^2)\n
  &\qquad\qquad
  +16(-245+968g-972g^2+224g^3)z+4(805-1512g+492g^2)z^2\n
  &\qquad\qquad
  +20(22g-35)z^3+35z^4\bigr),\\
  \mathcal{D}&=\{1^{\I},3^{\I}\}:\n
  &R_{-1}(z)=24576\bigl(8(1+2g)(5+2g)(17+6g)(39+10g)\n
  &\qquad\qquad
  +16(1625+2984g+1500g^2+224g^3)z+4(2005+2136g+492g^2)z^2\n
  &\qquad\qquad
  +20(47+22g)z^3+35z^4\bigr),\\
  \mathcal{D}&=\{1^{\II},3^{\II}\}:\n
  &R_{-1}(z)=24576\bigl(8(2g-7)(2g-3)(6g-7)(10g-1)\n
  &\qquad\qquad
  +16(-647+1736g-1188g^2+224g^3)z+4(1333-1800g+492g^2)z^2\n
  &\qquad\qquad
  +20(22g-41)z^3+35z^4\bigr),\\
  \mathcal{D}&=\{1^{\I},2^{\I},3^{\I}\}:\n
  &R_{-1}(z)=12288\bigl(24(1+2g)(7+2g)(251+144g+20g^2)\n
  &\qquad\qquad
  +16(2411+3944g+1716g^2+224g^3)z+4(2629+2424g+492g^2)z^2\n
  &\qquad\qquad
  +20(53+22g)z^3+35z^4\bigr),\\
  \mathcal{D}&=\{1^{\II},2^{\II},3^{\II}\}:\n
  &R_{-1}(z)=12288\bigl(8(2g-7)(2g-5)(2g-3)(30g-7)\n
  &\qquad\qquad
  +16(-1145+2552g-1404g^2+224g^3)z+4(1885-2088g+492g^2)z^2\n
  &\qquad\qquad
  +20(22g-47)z^3+35z^4\bigr),\\
  \mathcal{D}&=\{1^{\I},1^{\II}\}:\n
  &R_{-1}(z)=73728\bigl(8(2g-3)(1+2g)(5+6g)(19+10g)\n
  &\qquad\qquad
  +16(-135-328g+156g^2+224g^3)z+4(-299+168g+492g^2)z^2\n
  &\qquad\qquad
  +20(3+22g)z^3+35z^4\bigr).
\end{align}

\bigskip

We have also calculated for $K=10$ and $K=12$ cases,
\begin{align*}
  K=10:\ \ \mathcal{D}&=\{4^{\I}\},\ \{4^{\II}\},\ \{1^{\I},4^{\I}\},
  \ \{1^{\II},4^{\II}\},\ \{2^{\I},3^{\I}\},\ \{2^{\II},3^{\II}\},
  \ \{1^{\I},2^{\I},4^{\I}\},\ \{1^{\II},2^{\II},4^{\II}\},\\
  &\quad\ \{1^{\I},2^{\I},3^{\I},4^{\I}\},
  \ \{1^{\II},2^{\II},3^{\II},4^{\II}\},\ \{2^{\I},1^{\II}\},
  \ \{1^{\I},2^{\II}\},\\
  K=12:\ \ \mathcal{D}&=\{5^{\I}\},\ \{5^{\II}\},
  \ \{1^{\I},5^{\I}\},\ \{1^{\II},5^{\II}\},\ \{2^{\I},4^{\I}\},
  \ \{2^{\II},4^{\II}\},\ \{1^{\I},2^{\I},5^{\I}\},
  \ \{1^{\II},2^{\II},5^{\II}\},\\
  &\quad\ \{1^{\I},3^{\I},4^{\I}\},\ \{1^{\II},3^{\II},4^{\II}\},
  \ \{1^{\I},2^{\I},3^{\I},5^{\I}\},\ \{1^{\II},2^{\II},3^{\II},5^{\II}\},
  \ \{1^{\I},2^{\I},3^{\I},4^{\I},5^{\I}\},\\
  &\quad\ \{1^{\II},2^{\II},3^{\II},4^{\II},5^{\II}\},\ \{3^{\I},1^{\II}\},
  \ \{1^{\I},3^{\II}\},\ \{1^{\I},2^{\I},1^{\II}\},
  \ \{1^{\I},1^{\II},2^{\II}\},\ \{2^{\I},2^{\II}\},
\end{align*}
and checked Conjecture \ref{conj_gcr} and Conjecture \ref{conj_alphaj}.

\subsection{Jacobi type}
\label{app:J}

We present $R_{-1}(z)$ for all $\mathcal{D}$ with $K\leq 6$.

\noindent
\underline{$K=2$}
\begin{equation}
  \mathcal{D}=\{\}:
  \ \ R_{-1}(z)=16b(a-1).
\end{equation}
\underline{$K=4$}
\begin{align}
  \mathcal{D}&=\{1^{\I}\}:
  \ \ R_{-1}(z)=128(2+b)\Bigl(
  2(a-2)(a-1)\bigl((2+b)^2-3-a-2a^2\bigr)\n
  &\qquad\qquad\qquad\qquad\qquad\quad
  -\bigl((2+b)^2+1-10a+3a^2\bigr)z+z^2\Bigr),
  \label{J:1I}\\
  \mathcal{D}&=\{1^{\II}\}:
  \ \ R_{-1}(z)=\bigl(\text{$R_{-1}(z)$ in \eqref{J:1I}}\bigr)
  \bigl|_{b\to-b}.
\end{align}
\underline{$K=6$}
\begin{align}
  \mathcal{D}&=\{2^{\I}\}:\n
  &R_{-1}(z)=3072(1-a)\Bigl(
  2(a-3)(a-2)\bigl(3(a-2)a(2+a)(3+a)\n
  &\qquad\qquad\quad
  -3(2+a)(5+a+a^3)(3+b)-(14-3a+3a^2)(3+b)^2\n
  &\qquad\qquad\quad
  +(10+3a+3a^2)(3+b)^3+2(3+b)^4-(3+b)^5\bigr)\n
  &\qquad\qquad
  +3\bigl(2(a-2)a(-13+2a^2)+(46-28a-2a^2+12a^3-3a^4)(3+b)\n
  &\qquad\qquad\quad
  -2(1-6a-a^2)(3+b)^2-2(1+4a)(3+b)^3-2(3+b)^4+(3+b)^5\bigr)z\n
  &\qquad\qquad
  +6\bigl((a-2)a-(4-3a)(3+b)+2(3+b)^2-(3+b)^3\bigr)z^2\n
  &\qquad\qquad
  +3(3+b)z^3\Bigr),
  \label{J:2I}\\
  \mathcal{D}&=\{2^{\II}\}:
  \ \ R_{-1}(z)=-\bigl(\text{$R_{-1}(z)$ in \eqref{J:2I}}\bigr)
  \bigl|_{b\to-b},\\
  \mathcal{D}&=\{1^{\I},2^{\I}\}:\n
  &R_{-1}(z)=-1536(1-a)(4+b)\Bigl(
  2(a-3)(a-2)\bigl(3(a-2)a(2+a)(3+a)\n
  &\qquad\qquad\quad
  +3(2+a)(5+a+a^3)(3+b)-(14-3a+3a^2)(3+b)^2\n
  &\qquad\qquad\quad
  -(10+3a+3a^2)(3+b)^3+2(3+b)^4+(3+b)^5\bigr)\n
  &\qquad\qquad
  +3\bigl(2(a-2)a(-13+2a^2)-(46-28a-2a^2+12a^3-3a^4)(3+b)\n
  &\qquad\qquad\quad
  +2(1-6a-a^2)(3+b)^2+2(1+4a)(3+b)^3-2(3+b)^4-(3+b)^5\bigr)z\n
  &\qquad\qquad
  +6\bigl((a-2)a+(4-3a)(3+b)+2(3+b)^2+(3+b)^3\bigr)z^2\n
  &\qquad\qquad
  -3(3+b)z^3\Bigr),
  \label{J:1I2I}\\
  \mathcal{D}&=\{1^{\II},2^{\II}\}:
  \ \ R_{-1}(z)=-\bigl(\text{$R_{-1}(z)$ in \eqref{J:1I2I}}\bigr)
  \bigl|_{b\to-b}.
\end{align}

\subsection{Wilson type}
\label{app:W}

We present $R_{-1}(z)$ for all $\mathcal{D}$ with $K\leq 4$.

\noindent
\underline{$K=2$}
\begin{equation}
  \mathcal{D}=\{\}:
  \ \ R_{-1}(z)=-2z^2+(b_1-2b_2)z-(b_1-2)b_3.
\end{equation}
\underline{$K=4$}\quad
$\mathcal{D}=\{1^{\I}\}$ :
\begin{align}
  &R_{-1}(z)\n
  =&\frac18(b_1-3)(b_1-2)\Bigl(
  2b_1+b_1(\sigma_1-5\sigma_1^{\prime})
  -b_1(\sigma_1^2+7\sigma_1^{\prime\,2})
  +16(\sigma_1\sigma_2^{\prime}+4\sigma_1^{\prime}\sigma_2
  -3\sigma_1^{\prime}\sigma_2^{\prime})\n
  &\ \ +8\bigl(\sigma_1^2\sigma_2^{\prime}+\sigma_1\sigma_1^{\prime}b_1^2
  +\sigma_1\sigma_1^{\prime}(\sigma_2+15\sigma_2^{\prime})
  +\sigma_1^{\prime\,2}(21\sigma_2-6\sigma_2^{\prime})\bigr)
  -8\bigl(\sigma_1^3\sigma_2^{\prime}+\sigma_1^2\sigma_1^{\prime}
  (7\sigma_2+2\sigma_2^{\prime})\n
  &\ \ +\sigma_1\sigma_1^{\prime\,2}(14\sigma_2-15\sigma_2^{\prime})
  -9\sigma_1^{\prime\,3}\sigma_2
  -4\sigma_1\sigma_2^{\prime\,2}+4\sigma_1^{\prime}
  (2\sigma_2^2-4\sigma_2\sigma_2^{\prime}+\sigma_2^{\prime\,2})\bigr)\n
  &\ \ -16\bigl(\sigma_1\sigma_1^{\prime}(\sigma_1+3\sigma_1^{\prime})
  (\sigma_1\sigma_2^{\prime}+\sigma_1^{\prime}\sigma_2)
  -\sigma_1^2\sigma_2^{\prime\,2}-\sigma_1\sigma_1^{\prime}
  (5\sigma_2^2-8\sigma_2\sigma_2^{\prime}+7\sigma_2^{\prime\,2})
  +\sigma_1^{\prime\,2}\sigma_2(\sigma_2-4\sigma_2^{\prime})\bigr)\n
  &\ \ -16(\sigma_1\sigma_2^{\prime}+\sigma_1^{\prime}\sigma_2)
  \bigl(\sigma_1^2\sigma_2^{\prime}-3\sigma_1\sigma_1^{\prime}
  (\sigma_2-\sigma_2^{\prime})-\sigma_1^{\prime\,2}\sigma_2\bigr)\Bigr)\n
  &+\frac12\Bigl(-6+9(3\sigma_1+5\sigma_1^{\prime})
  +8(\sigma_1^2+2\sigma_1\sigma_1^{\prime}+16\sigma_1^{\prime\,2}
  -12\sigma_2+6\sigma_2^{\prime})
  -2(13\sigma_1^3+53\sigma_1^2\sigma_1^{\prime}
  +143\sigma_1\sigma_1^{\prime\,2}\n
  &\ \ -17\sigma_1^{\prime\,3}
  +52\sigma_1\sigma_2^{\prime}+220\sigma_1^{\prime}\sigma_2)
  +8\bigl(\sigma_1^4+7\sigma_1^3\sigma_1^{\prime}
  +15\sigma_1^2\sigma_1^{\prime\,2}-17\sigma_1\sigma_1^{\prime\,3}
  -2\sigma_1^{\prime\,4}
  +\sigma_1^2(9\sigma_2+11\sigma_2^{\prime})\n
  &\ \ +2\sigma_1\sigma_1^{\prime}(37\sigma_2-15\sigma_2^{\prime})
  -\sigma_1^{\prime\,2}(11\sigma_2+13\sigma_2^{\prime})
  +12\sigma_2(\sigma_2-2\sigma_2^{\prime})\bigr)
  -8\bigl(\sigma_1\sigma_1^{\prime}
  (\sigma_1^3-4\sigma_1^2\sigma_1^{\prime}
  -16\sigma_1\sigma_1^{\prime\,2}\n
  &\ \ -5\sigma_1^{\prime\,3})
  +\sigma_1^3(3\sigma_2+5\sigma_2^{\prime})
  +\sigma_1^2\sigma_1^{\prime}(23\sigma_2-33\sigma_2^{\prime})
  -\sigma_1\sigma_1^{\prime\,2}(35\sigma_2+11\sigma_2^{\prime})
  -\sigma_1^{\prime\,3}(17\sigma_2-\sigma_2^{\prime})\n
  &\ \ +2\sigma_1
  (6\sigma_2^2-12\sigma_2\sigma_2^{\prime}
  +11\sigma_2^{\prime\,2})
  +2\sigma_1^{\prime}(\sigma_2^2+4\sigma_2\sigma_2^{\prime}
  +6\sigma_2^{\prime\,2})\bigr)
  -8\bigl(2\sigma_1^2\sigma_1^{\prime\,2}b_1^2
  -\sigma_1^4\sigma_2^{\prime}
  -\sigma_1^3\sigma_1^{\prime}(\sigma_2\n
  &\ \ -11\sigma_2^{\prime})
  +\sigma_1^2\sigma_1^{\prime\,2}(27\sigma_2-5\sigma_2^{\prime})
  +\sigma_1\sigma_1^{\prime\,3}(19\sigma_2-9\sigma_2^{\prime})
  -\sigma_1^{\prime\,4}\sigma_2
  -\sigma_1^2(3\sigma_2^2-6\sigma_2\sigma_2^{\prime}
  +19\sigma_2^{\prime\,2})\n
  &\ \ +4\sigma_1\sigma_1^{\prime}(3\sigma_2
  -5\sigma_2^{\prime})(\sigma_2+\sigma_2^{\prime})
  +\sigma_1^{\prime\,2}(\sigma_2-\sigma_2^{\prime})
  (9\sigma_2+7\sigma_2^{\prime})\bigr)
  +8\bigl(\sigma_1^4\sigma_1^{\prime}\sigma_2^{\prime}
  +\sigma_1^3\sigma_1^{\prime\,2}(5\sigma_2-4\sigma_2^{\prime})\n
  &\ \ +\sigma_1^2\sigma_1^{\prime\,3}(4\sigma_2
  -5\sigma_2^{\prime})
  -\sigma_1\sigma_1^{\prime\,4}\sigma_2
  -4\sigma_1^3\sigma_2^{\prime\,2}+\sigma_1^2\sigma_1^{\prime}
  (5\sigma_2-7\sigma_2^{\prime})(\sigma_2+\sigma_2^{\prime})\n
  &\ \ +\sigma_1\sigma_1^{\prime\,2}
  (7\sigma_2-5\sigma_2^{\prime})(\sigma_2+\sigma_2^{\prime})
  +4\sigma_1^{\prime\,3}\sigma_2^2\bigr)\Bigr)z\n
  &-2\Bigl(14+13\sigma_1+67\sigma_1^{\prime}
  -6(3\sigma_1^2+11\sigma_1\sigma_1^{\prime}-8\sigma_1^{\prime\,2}
  +10\sigma_2-2\sigma_2^{\prime})
  +2\bigl(2\sigma_1^3+3\sigma_1^2\sigma_1^{\prime}
  -48\sigma_1\sigma_1^{\prime\,2}\n
  &\ \ -\sigma_1^{\prime\,3}
  +\sigma_1(21\sigma_2-5\sigma_2^{\prime})-\sigma_1^{\prime}
  (31\sigma_2+9\sigma_2^{\prime})\bigr)
  +2\bigl(2\sigma_1\sigma_1^{\prime}(\sigma_1^2+12\sigma_1\sigma_1^{\prime}
  -2\sigma_1^{\prime\,2})
  -3\sigma_1^2(\sigma_2-\sigma_2^{\prime})\n
  &\ \ +6\sigma_1\sigma_1^{\prime}
  (7\sigma_2-\sigma_2^{\prime})+\sigma_1^{\prime\,2}
  (9\sigma_2-13\sigma_2^{\prime})
  +2(3\sigma_2^2-6\sigma_2\sigma_2^{\prime}-5\sigma_2^{\prime\,2})\bigr)
  -2\bigl(3\sigma_1^2\sigma_1^{\prime\,2}(\sigma_1-\sigma_1^{\prime})\n
  &\ \ +\sigma_1^3\sigma_2^{\prime}
  -\sigma_1^{\prime\,3}\sigma_2
  +\sigma_1^2\sigma_1^{\prime}(11\sigma_2-6\sigma_2^{\prime})
  +\sigma_1\sigma_1^{\prime\,2}(6\sigma_2-11\sigma_2^{\prime})
  +\sigma_1(3\sigma_2^2-6\sigma_2\sigma_2^{\prime}-5\sigma_2^{\prime\,2})\n
  &\ \ +\sigma_1^{\prime}(5\sigma_2^2
  +6\sigma_2\sigma_2^{\prime}
  -3\sigma_2^{\prime\,2})\bigr)\Bigr)z^2\n
  &+8\bigl(8-2(2\sigma_1-7\sigma_1^{\prime})
  -(13\sigma_1-3\sigma_1^{\prime})\sigma_1^{\prime}-6\sigma_2
  +2\sigma_2^{\prime}
  +3(\sigma_1-\sigma_1^{\prime})\sigma_1\sigma_1^{\prime}
  +3(\sigma_1\sigma_2-\sigma_1^{\prime}\sigma_2^{\prime})\n
  &\ \ -\sigma_1\sigma_2^{\prime}+\sigma_1^{\prime}\sigma_2\bigr)z^3\n
  &+12(-2+\sigma_1-\sigma_1^{\prime})z^4,
  \label{W:1I}\\
  &\!\!\!\mathcal{D}=\{1^{\II}\}:
  \ \ R_{-1}(z)=\bigl(\text{$R_{-1}(z)$ in \eqref{W:1I}}\bigr)
  \bigl|_{(a_1,a_2)\leftrightarrow(a_3,a_4)}.
\end{align}

\subsection{Askey-Wilson type}
\label{app:AW}

We present $R_{-1}(z)$ for all $\mathcal{D}$ with $K\leq 4$.

\noindent
\underline{$K=2$}
\begin{equation}
  \mathcal{D}=\{\}:
  \ \ R_{-1}(z)=-\frac12q^{-1}(q-1)^2
  \bigl((b_1+b_3q^{-1})z+(1-b_4q^{-2})(b_1-b_3)\bigr).
\end{equation}
\underline{$K=4$}\quad
$\mathcal{D}=\{1^{\I}\}$ :
\begin{align}
  &R_{-1}(z)\times\frac{2q^{\frac{17}{2}}\sigma_2}{(1-q)^4(1+q)}\n
  =&(q^2-\sigma_2\sigma_2^{\prime})(q^3-\sigma_2\sigma_2^{\prime})\Bigl(
  -q(1+q)(\sigma_1^2-q^2\sigma_1^{\prime\,2})
  -(1-q)(1-q-q^2)(\sigma_2-q^2\sigma_2^{\prime})\n
  &\ \ +\sigma_1^2\bigl(\sigma_2+q(1+q+q^2)\sigma_2^{\prime}\bigr)
  +q^2(1+q)\sigma_1\sigma_1^{\prime}(\sigma_2-q^2\sigma_2^{\prime})
  -q\sigma_1^{\prime\,2}\bigl((1+q+q^2)\sigma_2+q^3\sigma_2^{\prime}\bigr)\n
  &\ \ -(1-q)(\sigma_2^2-q^4\sigma_2^{\prime\,2})
  -(1+q)\bigl((1-q)\sigma_1^2\sigma_2^{\prime}
  (\sigma_2+q^3\sigma_2^{\prime})
  +\sigma_1\sigma_1^{\prime}(\sigma_2^2-q^4\sigma_2^{\prime\,2})\n
  &\ \ -q(1-q)\sigma_1^{\prime\,2}\sigma_2(\sigma_2+q\sigma_2^{\prime})
  -2(1-q)^2 \sigma_2\sigma_2^{\prime}(\sigma_2-q^2\sigma_2^{\prime})\bigr)
  -q^2\sigma_1^2\sigma_2^{\prime\,2}
  \bigl((1+q+q^2)\sigma_2+q^3\sigma_2^{\prime}\bigr)\n
  &\ \ +(1+q)\sigma_1\sigma_1^{\prime}\sigma_2\sigma_2^{\prime}
  (\sigma_2-q^2\sigma_2^{\prime})
  +q\sigma_1^{\prime\,2}\sigma_2^2
  \bigl(\sigma_2+q(1+q+q^2)\sigma_2^{\prime}\bigr)
  +(1-q)\sigma_2\sigma_2^{\prime}(\sigma_2^2-q^4\sigma_2^{\prime\,2})\n
  &\ \ +\sigma_2\sigma_2^{\prime}\bigl(q(1+q)
  (q^2\sigma_1^2\sigma_2^{\prime\,2}-\sigma_1^{\prime\,2}\sigma_2^2)
  -(1-q)(1+q-q^2)\sigma_2\sigma_2^{\prime}
  (\sigma_2-q^2\sigma_2^{\prime})\bigr)\Bigr)\n
  &+q\Bigl(
  q^4\bigl(-3q(1+q)\sigma_1^2+3q^3(1+q)\sigma_1^{\prime\,2}-(3-8q+3q^3)
  (\sigma_2-q^2\sigma_2^{\prime})\bigr)
  +q^3\Bigl(2q\sigma_1^2\bigl(\sigma_2\n
  &\ \ +q(1+q+q^2)\sigma_2^{\prime}\bigr)
  -(1+q)(1+q+q^2-2q^3)\sigma_1\sigma_1^{\prime}(\sigma_2-q^2\sigma_2^{\prime})
  -2q^2\sigma_1^{\prime\,2}\bigl((1+q+q^2)\sigma_2\n
  &\ \ +q^3\sigma_2^{\prime}\bigr)
  -(1+3q-2q^2)(\sigma_2^2-q^4\sigma_2^{\prime\,2})\Bigr)
  +q\Bigl(q(1+q)\sigma_1^2\sigma_2^{\prime}
  \bigl((1+2q)(2-2q+q^2)\sigma_2\n
  &\ \ -q^2(1+q)(1-2q+3q^2-q^3)\sigma_2^{\prime}\bigr)
  +q(1+q)(1+q-q^2)\sigma_1\sigma_1^{\prime}
  (\sigma_2^2-q^4\sigma_2^{\prime\,2})\n
  &\ \ +q(1+q)\sigma_1^{\prime\,2}\sigma_2
  \bigl((1+q)(1-2q+3q^2-q^3)\sigma_2-q^2(1+2q)(2-2q+q^2)\sigma_2^{\prime}\bigr)
  +(2-4q\n
  &\ \ -6q^2+13q^3-4q^4-6q^5+3q^6)\sigma_2\sigma_2^{\prime}
  (\sigma_2-q^2\sigma_2^{\prime})\Bigr)
  +q\Bigl(-\sigma_1^2\sigma_2^{\prime}
  \bigl((1+2q+q^3)\sigma_2^2\n
  &\ \ +2q^2(1-q^3)\sigma_2\sigma_2^{\prime}
  -q^4(1+2q^2+q^3)\sigma_2^{\prime\,2}\bigr)
  +(1+q)(1-q-2q^2-q^3+q^4)
  \sigma_1\sigma_1^{\prime}\sigma_2\sigma_2^{\prime}\n
  &\ \ \times(\sigma_2-q^2\sigma_2^{\prime})
  -\sigma_1^{\prime\,2}\sigma_2\bigl((1+2q^2+q^3)\sigma_2^2
  -2q^2(1-q^3)\sigma_2\sigma_2^{\prime}
  -q^4(1+2q+q^3)\sigma_2^{\prime\,2}\bigr)\n
  &\ \ +(1+q)(3-4q+3q^2)\sigma_2\sigma_2^{\prime}(\sigma_2^2
  -q^4\sigma_2^{\prime\,2})\Bigr)
  +\sigma_2\sigma_2^{\prime}\Bigl(
  -(1+q)\sigma_1^2\sigma_2^{\prime}\bigl((1+q)(1-3q+2q^2-q^3)\sigma_2\n
  &\ \ +q^3(2+q)(1-2q+2q^2)\sigma_2^{\prime}\bigr)
  -(1+q)(1-q-q^2)\sigma_1\sigma_1^{\prime}(\sigma_2^2-q^4\sigma_2^{\prime\,2})
  +q(1+q)\sigma_1^{\prime\,2}\sigma_2\n
  &\ \ \times\bigl((2+q)(1-2q+2q^2)\sigma_2
  +q(1+q)(1-3q+2q^2-q^3)\sigma_2^{\prime}\bigr)
  +(3-6q-4q^2+13q^3\n
  &\ \ -6q^4-4q^5+2q^6)\sigma_2\sigma_2^{\prime}
  (\sigma_2-q^2\sigma_2^{\prime})\Bigr)
  +\sigma_2\sigma_2^{\prime}\Bigl(
  -2q^2\sigma_1^2\sigma_2^{\prime\,2}
  \bigl((1+q+q^2)\sigma_2+q^3\sigma_2^{\prime}\bigr)\n
  &\ \ +(1+q)(2-q-q^2-q^3)\sigma_1\sigma_1^{\prime}\sigma_2
  \sigma_2^{\prime}(\sigma_2-q^2\sigma_2^{\prime})
  +2q\sigma_1^{\prime\,2}\sigma_2^2
  \bigl(\sigma_2+q(1+q+q^2)\sigma_2^{\prime}\bigr)\n
  &\ \ +(2-3q-q^2)\sigma_2\sigma_2^{\prime}(\sigma_2^2
  -q^4\sigma_2^{\prime\,2})\Bigr)
  +\sigma_2^2\sigma_2^{\prime\,2}
  \bigl(3q(1+q)(q^2\sigma_1^2\sigma_2^{\prime\,2}
  -\sigma_1^{\prime\,2}\sigma_2^2)\n
  &\ \ -(3-8q^2+3q^3)\sigma_2\sigma_2^{\prime}
  (\sigma_2-q^2\sigma_2^{\prime})\bigr)\Bigr)z\n
  &+q^2\Bigl(
  -3q^3\bigl(q(1+q)(\sigma_1^2-q^2\sigma_1^{\prime\,2})
  +(1-4q+q^3)(\sigma_2-q^2\sigma_2^{\prime})\bigr)
  +q^2\bigl(q \sigma_1^2\sigma_2+q^2(1+q+q^2)\sigma_1^2\sigma_2^{\prime}\n
  &\ \ -(1+q)(2+2q+2q^2-q^3)\sigma_1\sigma_1^{\prime}
  (\sigma_2-q^2\sigma_2^{\prime})
  -q^5 \sigma_1^{\prime\,2}\sigma_2^{\prime}
  -q^2(1+q+q^2)\sigma_1^{\prime\,2}\sigma_2\n
  &\ \ -(2+3q-q^2)(\sigma_2^2-q^4\sigma_2^{\prime\,2})\bigr)
  +q(1+q)\sigma_1^2\sigma_2^{\prime}
  \bigl((1+q-2q^2+q^3)\sigma_2-q^2(1-2q+q^2+q^3)\sigma_2^{\prime}\bigr)\n
  &\ \ +q(1+q)^2\sigma_1\sigma_1^{\prime}(\sigma_2^2-q^4\sigma_2^{\prime\,2})
  +q(1+q)\sigma_1^{\prime\,2}\sigma_2
  \bigl((1-2q+q^2+q^3)\sigma_2-q^2(1+q-2q^2+q^3)\sigma_2^{\prime}\bigr)\n
  &\ \ +(1-4q+q^3)(1-4q^2+q^3)\sigma_2\sigma_2^{\prime}
  (\sigma_2-q^2\sigma_2^{\prime})
  -q^2\sigma_1^2\sigma_2^{\prime\,2}
  \bigl((1+q+q^2)\sigma_2+q^3\sigma_2^{\prime}\bigr)\n
  &\ \ +(1+q)\bigl(1-2q(1+q+q^2)\bigr)
  \sigma_1\sigma_1^{\prime}\sigma_2\sigma_2^{\prime}
  (\sigma_2-q^2\sigma_2^{\prime})
  +q\sigma_1^{\prime\,2}\sigma_2^2
  \bigl(\sigma_2+q(1+q+q^2)\sigma_2^{\prime}\bigr)\n
  &\ \ +(1-3q-2q^2)\sigma_2\sigma_2^{\prime}
  (\sigma_2^2-q^4\sigma_2^{\prime\,2})
  +3\sigma_2\sigma_2^{\prime}
  \bigl(q^3(1+q)\sigma_1^2\sigma_2^{\prime\,2}
  -q(1+q)\sigma_1^{\prime\,2}\sigma_2^2\n
  &\ \ -(1-4q^2+q^3)\sigma_2\sigma_2^{\prime}
  (\sigma_2-q^2\sigma_2^{\prime})\bigr)\Bigr)z^2\n
  &+q^3\Bigl(
  -q^2\bigl(q(1+q)(\sigma_1^2-q^2\sigma_1^{\prime\,2})
  +(1-8q+q^3)(\sigma_2-q^2\sigma_2^{\prime})\bigr)
  -q(1+q)(\sigma_2-q^2\sigma_2^{\prime})\n
  &\ \ \times\bigl((1+q+q^2)\sigma_1\sigma_1^{\prime}
  +\sigma_2+q^2\sigma_2^{\prime}\bigr)
  +q(1+q)(q^2\sigma_1^2\sigma_2^{\prime\,2}-\sigma_1^{\prime\,2}\sigma_2^2)\n
  &\ \ -(1-8q^2+q^3)\sigma_2\sigma_2^{\prime}
  (\sigma_2-q^2\sigma_2^{\prime})\Bigr)z^3\n
  &+2q^6(\sigma_2-q^2\sigma_2^{\prime})z^4,
  \label{AW:1I}\\
  &\!\!\!\mathcal{D}=\{1^{\II}\}:
  \ \ R_{-1}(z)=\bigl(\text{$R_{-1}(z)$ in \eqref{AW:1I}}\bigr)
  \bigl|_{(a_1,a_2)\leftrightarrow(a_3,a_4)}.
\end{align}


\end{document}